\documentclass[english]{article}
\usepackage[T1]{fontenc}
\usepackage[latin9]{luainputenc}
\usepackage{geometry}
\usepackage{amstext}
\usepackage{amssymb}
\usepackage{feyn}
\usepackage{tikz}
\usepackage{graphicx}
\usepackage{array}
\usepackage{booktabs}

\usepackage{amsmath}
\usepackage{amsfonts}

\usepackage[normalem]{ulem}
\usepackage{color}
\usepackage{listings,braket}
\definecolor{darkgreen}{rgb}{0,0.35,0}
\definecolor{Rood}{rgb}{1, 0, 0}

\makeatletter
\usepackage[english]{babel}
\usepackage{feyn}

\makeatother

\begin{document}

\title{\textbf{The Abelian Higgs model under a gauge invariant looking glass: exploiting new Ward identities for gauge invariant operators and the Equivalence Theorem}}

{\author{\textbf{D.~Dudal$^{1,2}$}\thanks{david.dudal@kuleuven.be},
\textbf{G.~Peruzzo$^3$}\thanks{gperuzzofisica@gmail.com},
\textbf{S.~P.~Sorella$^3$}\thanks{silvio.sorella@gmail.com},\\\\\
\textit{{\small $^1$ KU Leuven Campus Kortrijk---Kulak, Department of Physics, Etienne Sabbelaan 53 bus 7657, 8500 Kortrijk, Belgium,}} \\
\textit{{\small $^2$ Ghent University, Department of Physics and Astronomy, Krijgslaan 281-S9, 9000 Gent, Belgium}}\\
\textit{{\small $^3$UERJ -- Universidade do Estado do Rio de Janeiro,}}\\
\textit{{\small Instituto de F\'{\i}sica -- Departamento de F\'{\i}sica Te\'orica -- Rua S\~ao Francisco Xavier 524,}}\\
\textit{{\small 20550-013, Maracan\~a, Rio de Janeiro, Brasil}}\\
}

\date{}

\maketitle
\begin{abstract}
The content of two additional Ward identities exhibited by the $U(1)$ Higgs model is exploited. These novel Ward identities can be derived only when a pair of local composite operators providing a gauge invariant setup for the Higgs particle and the massive vector boson is introduced in the theory from the beginning. Among the results obtained from the above mentioned Ward identities, we underline a new exact relationship between the stationary condition for the vacuum energy, the vanishing of the tadpoles and the vacuum expectation value of the gauge invariant scalar operator. We also present a characterization of the two-point correlation function of the composite operator corresponding to the vector boson in terms of the two-point function of the elementary gauge fields. Finally, a discussion on the connection between the cartesian and the polar parametrization of the complex scalar field is presented in the light of the Equivalence Theorem. The latter can in the current case be understood in the language of a constrained cohomology, which also allows to rewrite the action in terms of the aforementioned gauge invariant operators. We also comment on the diminished role of the global $U(1)$ symmetry and its breaking.
\end{abstract}

\section{Introduction}

In a series of previous works \cite{Dudal:2019aew,Dudal:2019pyg,Capri:2020ppe} the $U(1)$ Higgs model has been analyzed  within the gauge invariant formulation outlined in \cite{hooft2012nonperturbative,Frohlich:1980gj,Frohlich:1981yi}, see also \cite{Maas:2019nso,Maas:2017xzh,Sondenheimer:2019idq} for recent contributions. More precisely, the field content of the theory enables one to introduce the following pair of local gauge invariant composite operators $(O(x), V_\mu(x))$:
\begin{eqnarray}
O\left(x\right) & = & \frac{1}{2}\left(h^{2}+2vh+\rho^{2}\right) \,, \nonumber \\
V_{\mu}\left(x\right) & = & \frac{1}{2}\left(-\rho\partial_{\mu}h+h\partial_{\mu}\rho+v\partial_{\mu}\rho+eA_{\mu}\left(v^{2}+h^{2}+2vh+\rho^{2}\right)\right) \,, \label{ovop}
\end{eqnarray}
where $(h,\rho)$ stand for the Higgs and Goldstone fields, while $v$ is the classical minimum of the Higgs potential, eqs.~(\ref{ha}), (\ref{exp}). For the record, we will work in Euclidean convention in this paper.

In particular, from the one-loop computation of the two point functions $\langle O(p)O(-p) \rangle$, $\langle V_\mu(p)V_\nu(-p) \rangle^T$ carried out in \cite{Dudal:2019pyg}
in the 't Hooft $R_\xi$ gauge, it turns out that, besides being independent from the gauge parameter $\xi$,
the pole masses of $\langle O(p)O(-p) \rangle$, $\langle V_\mu(p)V_\nu(-p) \rangle^T$ coincide, respectively, with the pole masses of the corresponding elementary correlation functions $\langle h(p)h(-p) \rangle$  and
$\langle A_\mu(p)A_\nu(-p) \rangle^T$, where $T$ denotes the transverse
components\footnote{The correlation function $\langle V_\mu(p) V_\nu(-p) \rangle $ can be decomposed into transverse and longitudinal components, according to
\begin{equation}
\langle V_\mu(p) V_\nu(-p) \rangle = V^T(p^2) {\cal P}_{\mu\nu}(p) + V^L(p^2) {\cal L}_{\mu\nu}(p) \;, \label{PL}
\end{equation}
where ${\cal P}_{\mu\nu}$ and ${\cal L}_{\mu\nu}$ are the transverse and longitudinal projectors, namely
\begin{equation}
{\cal P}_{\mu\nu}(p) = \left( \delta_{\mu\nu} - \frac{p_\mu p_\nu}{p^2} \right) \;, \qquad {\cal L}_{\mu\nu}(p) = \frac{p_\mu p_\nu}{p^2} \;. \label{pj}
\end{equation} An analogue decomposition can be introduced for $\langle A_\mu(p) A_\nu(-p)\rangle$. Moreover, in the Landau gauge, $ \partial A=0$, the two-point function $\langle A_\mu(p) A_\nu(-p)\rangle$ is already transverse.
} of $\langle V_\mu(p)V_\nu(-p) \rangle$ and $ \langle A_\mu(p)A_\nu(-p) \rangle$. Also, both tree-level and one-loop expressions of the longitudinal part of $\langle V_\mu(p)V_\nu(-p) \rangle^L$ remain independent from the momentum $p^2$, so that they are not associated
to any physical mode. Finally, both correlation functions $\langle O(p)O(-p) \rangle$, $\langle V_\mu(p)V_\nu(-p) \rangle^T$ display a K{\"a}ll{\'e}n-Lehmann (KL) representation with positive and gauge parameter independent spectral densities \cite{Dudal:2019pyg}, a feature  which highlights the relevance of the operators $(O(x), V_\mu(x))$ in order to  provide  a local and renormalizable framework \cite{Capri:2020ppe} for a fully gauge invariant description of the Higgs and vector gauge bosons.

Following the standard quantum field theory setup \cite{Itzykson:1980rh}, in order to study the correlation functions of the  local composite operators $(O(x), V_\mu(x))$ one has to introduce them in the starting action by means of a suitable pair of external sources $(J(x), \Omega_\mu(x))$. It is remarkable that, in doing so, two novel powerful Ward identities \cite{Capri:2020ppe}, see eqs.~(\ref{hequation}), (\ref{Widom}), arise which enable us to establish a set of non-trivial all orders statements about $(O(x), V_\mu(x))$. For instance, in \cite{Capri:2020ppe}, these Ward identities have been already employed  to prove the all order renormalizability of $(O(x), V_\mu(x))$ as well as to detect their possible mixing with other gauge invariant composite operators. Moreover, besides the renormalization aspects, these Ward identities encode many other features whose description is the main aim of this paper. Let us also underline that the aforementioned Ward identities can be obtained only when the operators $(O(x), V_\mu(x))$ are manifestly present in the starting action, yielding a further striking evidence of the role played by  these operators in the gauge invariant picture of the Higgs model.

For the benefit of the reader, we enlist here the set of results which give rise to the content of the present work:
\begin{itemize}
\item After providing a short self-contained summary, Section~\ref{S2},  of the quantization of the $U(1)$ Higgs model in the Landau gauge\footnote{Although other renormalizable covariant gauges, as the 't Hooft $R_\xi$ gauge, could be employed in the quantization of the model \cite{Dudal:2019aew,Dudal:2019pyg}, the gauge invariance of both operators $(O(x), V_\mu(x))$ strongly motivates  the  adoption of the Landau gauge as the most natural and direct choice \cite{Capri:2020ppe}.}, of the introduction of the local gauge invariant composite operators $(O(x), V_\mu(x))$ and of the respective Ward identities, we start by addressing the issue of the nature  of the vector operator $V_\mu(x)$, eq.~(\ref{ovop}), a topic covered in Section~\ref{S3}. It turns out that the vector operator $V_\mu$ is nothing but the conserved Noether current corresponding to the global $U(1)$ invariance of the Higgs model. Let us also point out that this global symmetry is manifestly preserved by the Landau gauge, being broken in the $R_\xi$ gauge by unphysical BRST exact terms which can be kept under control to all orders by means of the introduction of a suitable set of BRST doublets of external sources, see the extensive analysis worked out in \cite{Kraus:1995jk,Haussling:1996rq}. The conservation law of the vector current, through the  Ward identity (\ref{Widom}), is at origin of the all orders (non)-renormalization properties of $V_\mu(x)$,  as expressed by \cite{Capri:2020ppe}
\begin{eqnarray}
Z_{\Omega \Omega} & = & 1 \;, \qquad \Omega_{0\mu} = Z_{\Omega \Omega}\; \Omega_\mu \;, \nonumber \\
(V_0)_\mu &=& Z^{1/2}_h V_\mu \;, \qquad  h_0= Z^{1/2}_{h} h \;, \qquad Z_v = Z_h = Z_\rho \;, \nonumber \\
Z_e & = & Z^{-1/2}_A \;, \label{rz}
\end{eqnarray}
where $Z_{\Omega \Omega}$ stands for the renormalization factor of the external source $\Omega_\mu$ coupled to $V_\mu$ and $(V_0)_\mu $ denotes the bare operator written in terms of bare fields and bare parameters. In particular, as a consequence of $Z_{\Omega \Omega}=1$, the operator $V_\mu$ displays vanishing anomalous dimension, an expected feature for a conserved current \cite{Collins:2005nj}. Also, the last equation in (\ref{rz}) expresses a well known general feature of $U(1)$ gauge models, see for example  \cite{Itzykson:1980rh}.
\item We move then to the investigation of the exact consequences of the first Ward identity, eq.~(\ref{hequation}), directly related to the composite operator $O(x)$. As we shall see in details in Sections~\ref{VC} and \ref{OC}, this Ward identity enables us to establish  an exact bridge
between the BRST invariant counterterm accounting for the vanishing of the tadpole diagrams ($\left\langle h \right\rangle = 0$) which arise from one-loop onwards, another independent BRST invariant vacuum counterterm connected to the vacuum energy ${\cal E}_{v}$ and the vacuum expectation value of $\braket{O}$, namely
\begin{equation}
 \braket{h} - \frac{\partial {\cal E}_{v} }{\partial v} = \lambda v\braket{ O}\;. \label{Wennw}
\end{equation}
Let us remind here that the appearance of a BRST invariant counterterm from one-loop onwards allowing to kill tadpoles order per order, see the $\delta \sigma$ term in the fifth line of equation (\ref{fctt}), is a well known feature of the Higgs model since its original BRST formulation \cite{Becchi:1974md,Becchi:1974xu}, see also \cite{Kraus:1995jk,Haussling:1996rq} for more recent investigations. However, to our knowledge, it is the first time that such explicit relationship, encoded in \eqref{Wennw}, is  derived from a renormalizable Ward identity related to the construction of a gauge invariant framework for the Higgs model. 

\item Section~\ref{VV} is entirely devoted to the study of the two-point correlation function $\langle V_\mu(p) V_\nu(-p)\rangle $ by exploiting the second Ward identity, eq.~(\ref{Widom}), from which a couple of exact non-trivial statements about $\langle V_\mu(p) V_\nu(-p)\rangle$ can be derived. The main results are summarized in eqs.~(\ref{eq:transversal_result}), (\ref{eq:longitudinal_result}), which we report below, namely:
\begin{eqnarray}
\mathcal{P}_{\mu\nu}(p) \left\langle V_{\mu}\left(p\right)V_{\nu}\left(-p\right)\right\rangle & = & \frac{p^{4}}{4e^{2}}\mathcal{P}_{\mu\nu}\left(p\right)\left\langle A_{\mu}\left(p\right)A_{\nu}\left(-p\right)\right\rangle -3\frac{\left(p^{2}-m^{2}\right)}{4e^{2}}\label{ta1} \;,
\end{eqnarray}
and
\begin{eqnarray}
\mathcal{L}_{\mu\nu}(p)\left\langle V_{\mu}\left(p\right)V_{\nu}\left(-p\right)\right\rangle  & = & \frac{v^{2}}{4}\label{la2} \;,
\end{eqnarray}
where $\mathcal{P}_{\mu\nu}(p)$ and $\mathcal{L}_{\mu\nu}(p)$ are the transverse and longitudinal projectors, eq.~(\ref{pj}). From eq.~(\ref{ta1}) one can see that the transverse component of $\langle V_\mu(p) V_\nu(-p)\rangle $ is fully determined by the elementary two-point function of the massive gauge field $\langle A_\mu(p) A_\nu(-p)\rangle $. This result has a deep physical meaning, implying in fact that the pole masses of the transverse component of $\langle V_\mu(p) V_\nu(-p)\rangle $ and of $\langle A_\mu(p) A_\nu(-p)\rangle $ are identical. On the other hand, eq.~(\ref{la2}) states that the longitudinal component of $\langle V_\mu(p) V_\nu(-p)\rangle $ does not receive any quantum correction beyond the tree level one, which is moreover completely independent from the momentum $p^2$. Both results (\ref{ta1}), (\ref{la2}) confirm the usefulness of the conserved current operator $V_\mu$ for a genuine gauge invariant description of the gauge massive particle. In Appendix~\ref{A} one finds the details of the explicit one-loop verification of eqs.~(\ref{ta1}), (\ref{la2}).
\item Finally, in Section~\ref{EQ} we present an  account of the connection between the Higgs model expressed in {\it cartesian coordinates} and {\it polar coordinates}\footnote{With the name {\it cartesian coordinates} we refer to the parametrization of the complex scalar field $\varphi$ written as, see eq.~(\ref{exp}),
\begin{eqnarray}
\varphi_{\rm cart} & = & \frac{1}{\sqrt{2}}\left(v+h+i\rho\right)\;, \label{cart}
\end{eqnarray}
while, for {\it polar coordinates}, we refer to
\begin{eqnarray}
\varphi_{\rm pol} & = & \frac{1}{\sqrt{2}}\left(v+h'\right)\;e^{i\rho'} \;. \label{pol}
\end{eqnarray}
} in the light of the Equivalence Theorem \cite{Bergere:1975tr,Haag:1958vt,Kamefuchi:1961sb,Lam:1973qa}, a general result in quantum field theory stating that field redefinitions have no effects on physical observable quantities like the $S$-matrix amplitudes. Here, we shall follow the more recent approach outlined in \cite{Blasi:1998ph} where the Equivalence Theorem has been re-formulated within a BRST framework by exponentiating the Jacobian arising from the field redefinition {\it \`a la Faddeev-Popov}. In fact, when exponentiated, the Jacobian can be re-expressed by introducing a new set of harmless ghost variables, in much the same way as the Faddeev-Popov determinant. Further, BRST transformations leaving the transformed action invariant can be established for the new ghosts. As pointed out in \cite{Blasi:1998ph}, the new ghost fields compensate precisely the effects of the field redefinition. The existence of such BRST transformations guarantees that physical observables are insensitive to field redefinitions, according to the Equivalence Theorem. We will discuss this from the viewpoint of a constrained BRST cohomology. We will also briefly sketch how this Equivalence Theorem can also be used to rewrite the action in terms of the gauge invariant field operators \eqref{ovop}.
\end{itemize}

\section{Summary of the $U(1)$ Higgs model in the Landau gauge: introduction of the gauge invariant composite operators $O(x)$ and $V_\mu(x)$ and related Ward identities}\label{S2}
The aim of this section is that of providing a short self-consistent summary of the results obtained in  \cite{Dudal:2019aew,Dudal:2019pyg,Capri:2020ppe} when the composite gauge invariant operators $(O(x),V_\mu(x))$ are introduced from the beginning in the starting action.
\subsection{The $U(1)$ Higgs model in the Landau gauge}

The  Abelian $U(1)$ Higgs model \cite{Higgs:1964pj,Higgs:1964ia,Englert:1964et,Guralnik:1964eu} is described by the following
action

\begin{eqnarray}
S_{\rm Higgs} & = & \int d^{4}x\left[\frac{1}{4}F_{\mu\nu}F_{\mu\nu}+\left(D_{\mu}\varphi\right)^{\ast}\left(D_{\mu}\varphi\right)+\frac{1}{2}\lambda\left(\left|\varphi\right|^{2}-\frac{v^{2}}{2}\right)^{2}\right] \;, \label{ha}
\end{eqnarray}
where
\begin{eqnarray}
F_{\mu\nu} & = & \partial_{\mu}A_{\nu}-\partial_{\nu}A_{\mu} \;, \nonumber \\
D_{\mu}\varphi & = & (\partial_{\mu}+ieA_{\mu})\varphi \;, \label{def}
\end{eqnarray}
with $\varphi$ being a complex scalar field, $e$  the electric charge and  $\lambda$ the quartic self-coupling.

Expanding the complex field $\varphi$ around the minimum of the classical potential in eq.~(\ref{ha}), i.e.
\begin{eqnarray}
\varphi & = & \frac{1}{\sqrt{2}}\left(v+h+i\rho\right)\;, \label{exp}
\end{eqnarray}
where $h$ and $\rho$ are the Higgs  and the Goldstone fields, expression (\ref{ha}) becomes
\begin{eqnarray}
S_{\rm Higgs} & = & \int d^{4}x\left[\frac{1}{4}F_{\mu\nu}F_{\mu\nu}+\frac{1}{2}\partial_{\mu}h\partial_{\mu}h+\frac{1}{2}\partial_{\mu}\rho\partial_{\mu}\rho\right.\nonumber
  +\frac{1}{2}e^{2}v^{2}A_{\mu}A_{\mu}+evA_{\mu}\partial_{\mu}\rho+\frac{1}{2}\lambda v^{2}h^{2}\nonumber \\
 &  & \;\;\;\;\;-eA_{\mu}\rho\partial_{\mu}h+eA_{\mu}h\partial_{\mu}\rho+e^{2}vhA_{\mu}A_{\mu}\nonumber  +\frac{1}{2}e^{2}\rho^{2}A_{\mu}A_{\mu}+\frac{1}{2}e^{2}h^{2}A_{\mu}A_{\mu}\nonumber \\
 &  & \;\;\;\;\;\left.+\frac{1}{8}\lambda h^{4}+\frac{1}{8}\lambda\rho^{4}+\frac{1}{2}\lambda vh^{3}+\frac{1}{2}\lambda vh\rho^{2}+\frac{1}{4}\lambda h^{2}\rho^{2}\right] \;, \label{ah2}
\end{eqnarray}
showing that both gauge and Higgs fields have acquired a mass, given respectively by
\begin{equation}
m^2 =e^2 v^2  \;, \qquad m^2_h = \lambda v^2 \;. \label{mah}
\end{equation}
The field $\rho$, the {\it would-be Goldstone boson}, remains massless. The action (\ref{ah2}) is left invariant by the local gauge transformations
\begin{equation}
\delta_\alpha A_{\mu}  =  -\partial_{\mu}\alpha \;, \qquad
\delta_\alpha h =  -e\alpha \rho \;, \qquad \delta_\alpha \rho  =  e\alpha \left(v+h\right) \;, \label{omg}
\end{equation}
with $\alpha(x)$ a local gauge parameter:
\begin{equation}
\delta_\alpha S_{\rm Higgs} = 0 \;. \label{ahinv}
\end{equation}
In order to quantize the model, we employ the Landau gauge \cite{Clark:1974eq}, $\partial_\mu A_\mu=0$. Following the BRST procedure \cite{Becchi:1974md,Becchi:1974xu}, for the Landau gauge-fixing term we have
\begin{eqnarray}
S_{\rm gf}  =  \int d^{4}x\left(ib\partial_{\mu}A_{\mu}+\overline{c}\partial^{2}c\right) \;, \label{lgfx}
\end{eqnarray}
where $b$ stands for the Nakanishi-Lautrup field, while $c$ and $\overline{c}$ are the Faddeev-Popov ghosts. The local gauge invariance, eq.~(\ref{ahinv}), is now replaced by the exact nilpotent BRST invariance, namely
\begin{equation}
s\left(S_{\rm Higgs}+S_{\rm gf}\right)=0 \;,  \label{brsti}
\end{equation}
where
\begin{eqnarray}
sA_{\mu} & = & -\partial_{\mu}c \;, \qquad sc  =  0 \;, \nonumber \\
sh & = & -ec\rho \;, \qquad
s\rho  =  ec\left(v+h\right) \;, \nonumber \\
s\overline{c} & = & ib \;, \qquad sb  =  0\;, \nonumber \\
s^2 & = & 0 \;.  \label{ss}
\end{eqnarray}
Besides the BRST invariance, the action $\left(S_{\rm Higgs}+S_{\rm gf}\right)$ enjoys the discrete charge conjugation symmetry
\begin{eqnarray}
A_{\mu} & \rightarrow & -A_{\mu} \;, \qquad  h  \rightarrow  h \; \nonumber \\
\rho & \rightarrow & -\rho \;, \qquad b  \rightarrow  -b \;, \nonumber \\
\overline{c} & \rightarrow &  -\overline{c} \;, \qquad   c \rightarrow  -c \;, \label{cgj}
\end{eqnarray}
as well as the  global invariance
\begin{eqnarray}
\delta_{\omega}h & = & -e\omega\rho \;, \qquad  \delta_{\omega}\rho =e\omega\left(v+h\right)
\;, \nonumber \\
\delta_\omega A_\mu & = & 0\;, \qquad \delta_\omega \overline{c}=0  \;, \qquad \delta_\omega c = 0\;, \qquad \delta_\omega b= 0  \;, \label{gbo}
\end{eqnarray}
with
\begin{equation}
\delta_{\omega}  \left(S_{\rm Higgs}+S_{\rm gf}\right)  =  0 \;, \label{cnst}
\end{equation}
where $\omega$ is a constant parameter. As we shall see in the following, the global invariance, eq.~(\ref{cnst}), can be converted into a Ward identity which will imply helpful relationships between the various terms of the most general local invariant counterterm  needed to renormalize the model. It is worth observing here that, unlike the $R_\xi$ gauge \cite{Kraus:1995jk,Haussling:1996rq}, the Faddeev-Popov ghosts
$(\overline{c}, c)$ are now non-interacting fields, being completely decoupled. This is a helpful feature of the Landau gauge. The same property holds for the $b$-field, which appears only at the quadratic level.

Let us end this short summary by noticing that the composite operators $(O(x),V_{\mu}(x))$ are, respectively, even and odd under charge conjugation, i.e.
\begin{eqnarray}
O(x) \ & \rightarrow O(x) \;,
\nonumber \\
V_\mu(x) & \rightarrow - V_\mu(x) \;,  \label{odev}
\end{eqnarray}
a property which will be exploited in the next section.

\subsection{Introduction of the gauge  invariant operators $( O\left(x\right),V_{\mu}\left(x\right))$ and Ward identities}

Following \cite{Capri:2020ppe}, the gauge invariant operators $( O\left(x\right),V_{\mu}\left(x\right))$ can be studied at the quantum level by coupling them to  a pair of BRST invariant external sources $(J, \Omega_\mu)$. Moreover, taking into account the mixing between $V_\mu$ and the BRST invariant quantities $\partial_\nu F_{\nu\mu}$ and $\partial_\mu b$ as well as that of the scalar operator $O$ with $v^2$, for the complete starting action $\Sigma$ one needs  \cite{Capri:2020ppe}
\begin{eqnarray}
\Sigma & = & S_{\rm Higgs}+S_{\rm gf}+S_{\rm ext} \;, \label{cact}
\end{eqnarray}
with
\begin{equation}
S_{\rm ext}  =  \int d^{4}x \left(L(sh)+R(s\rho)  + JO+\eta v^{2}+\Omega_{\mu}V_{\mu}+\Upsilon_{\mu}\partial_{\nu}F_{\nu\mu}+\Theta_{\mu}\partial_{\mu}b \right) \;, \label{ext}
\end{equation}
where the sources $(L,R)$ allow to define the  BRST variations of the fields $(h, \rho)$ \cite{Piguet:1995er}, eqs.~(\ref{ss}), while $(J,\eta,\Omega_{\mu},\Upsilon_{\mu},\Theta_{\mu})$ couple, respectively,  to $(O(x), v^2, V_\mu(x), \partial_{\nu}F_{\nu\mu}, \partial_\mu b)$.  All external sources are BRST invariant, i.e.
\begin{equation}
sL= sR = sJ=s\eta=s\Omega_{\mu}=s\Upsilon_{\mu}=s\Theta_{\mu}=0 \;, \label{sext}
\end{equation}
so that $\Sigma$ is BRST invariant as well
\begin{equation}
s \Sigma = 0 \;. \label{ssig}
\end{equation}
The fields $(A_\mu,h, \rho, b)$ have dimensions $(1,1,1,2)$ and ghost number zero. The Faddeev-Popov ghosts $(\overline{c},c) $ have dimensions $(2,0)$ and ghost number $(-1,1)$. The two external sources $(L, R)$ have dimension $3$ and ghost number $-1$. Finally, the  sources $(J,\eta,\Omega_{\mu},\Upsilon_{\mu},\Theta_{\mu} )$ all have vanishing ghost number and dimensions $(2,2,1,1,1)$.

The complete classical action $\Sigma$ fulfills a huge number of Ward identities, which we enlist below:
\begin{itemize}
\item  the Slavnov-Taylor identity expressing the BRST invariance of $\Sigma$ at the functional level
\end{itemize}
\begin{eqnarray}
\mathcal{S}\left(\Sigma\right) & = & 0 \;, \label{slavnov}
\end{eqnarray}
where
\begin{eqnarray}
\mathcal{S}\left(\Sigma\right) & = & \int d^{4}x\left(-\partial_{\mu}c\frac{\delta\Sigma}{\delta A_{\mu}}+\frac{\delta\Sigma}{\delta L}\frac{\delta\Sigma}{\delta h}+\frac{\delta\Sigma}{\delta R}\frac{\delta\Sigma}{\delta\rho}+ib\frac{\delta\Sigma}{\delta\overline{c}}\right) \;. \label{st1}
\end{eqnarray}
\begin{itemize}
\item The $b$-Ward identity \cite{Piguet:1995er}
\end{itemize}
\begin{eqnarray}
\frac{\delta\Sigma}{\delta b} & = & i\partial_{\mu}A_{\mu}-\partial_{\mu}\Theta_{\mu}\;. \label{eq:bequation}
\end{eqnarray}
Notice that the right hand side of eq.~(\ref{eq:bequation}), being linear in the quantum fields, is a linear breaking, not affected by quantum corrections \cite{Piguet:1995er}. This equation expresses in functional form that the $b$ field is a non-interacting field.

\begin{itemize}
\item The antighost and ghost Ward identities
\end{itemize}
\begin{eqnarray}
\frac{\delta\Sigma}{\delta\overline{c}} & = & \partial^{2}c\label{eq:barcequation} \;,
\end{eqnarray}
and
\begin{eqnarray}
\frac{\delta\Sigma}{\delta c} & = & -\partial^{2}\overline{c}-Re\left(v+h\right)+Le\rho\label{eq:cequation} \;.
\end{eqnarray}
These two Ward identities express in functional form the decoupling of the Faddeev-Popov ghost fields in the Landau gauge.
\begin{itemize}
\item The global invariance, eq.~(\ref{gbo}), can be extended to the external sources in such a way that
\end{itemize}
\begin{eqnarray}
\delta_{\omega}h & = & -e\omega\rho \;, \qquad  \delta_{\omega}\rho =e\omega\left(v+h\right)
\;, \nonumber \\
\delta_\omega A_\mu & = & 0\;, \qquad \delta_\omega \overline{c}=0  \;, \qquad \delta_\omega c = 0\;, \qquad \delta_\omega b= 0  \;, \nonumber \\
\delta_{\omega}L & = & -e\omega R\;, \qquad \delta_{\omega}R  =  e\omega L \;, \nonumber \\
\delta_\omega J &= & \delta_\omega \eta= \delta_\omega \Omega_{\mu}= \delta_\omega \Upsilon_{\mu}= \delta_\omega \Theta_{\mu}=0 \;, \label{gbos}
\end{eqnarray}
with
\begin{equation}
\delta_{\omega}  \Sigma  =  0 \;, \label{cnstinv}
\end{equation}
yielding the powerful Ward identity
\begin{eqnarray}
\int d^{4}x\left[-\rho\frac{\delta\Sigma}{\delta h}+\left(v+h\right)\frac{\delta\Sigma}{\delta\rho}-R\frac{\delta\Sigma}{\delta L}+L\frac{\delta\Sigma}{\delta R}\right] & = & 0\label{eq:global} \;. \label{gboW}
\end{eqnarray}

\begin{itemize}
\item The charge conjugation  invariance
\end{itemize}
\begin{eqnarray}
A_{\mu} & \rightarrow & -A_{\mu} \;, \nonumber \\
h & \rightarrow & h\;, \nonumber \\
\rho & \rightarrow & -\rho \;, \nonumber \\
b & \rightarrow & -b \;, \nonumber \\
\overline{c} & \rightarrow & -\overline{c} \;, \nonumber \\
c & \rightarrow & -c \;, \nonumber \\
L & \rightarrow & L \;, \nonumber \\
R & \rightarrow & -R \;, \nonumber \\
J & \rightarrow & J \;, \nonumber \\
\Omega_{\mu} & \rightarrow & -\Omega_{\mu}\;, \nonumber \\
\Upsilon_{\mu} & \rightarrow & -\Upsilon_{\mu}\;, \nonumber \\
\Theta_{\mu} & \rightarrow & -\Theta_{\mu} \;. \label{wcc}
\end{eqnarray}
\begin{itemize}
\item The external sources Ward identities
\end{itemize}
\begin{eqnarray}
\frac{\delta\Sigma}{\delta\eta} & = & v^{2} \;, \nonumber \\
\frac{\delta\Sigma}{\delta\Upsilon_{\mu}} & = & \partial_{\nu}F_{\nu\mu} \;, \nonumber \\
\frac{\delta\Sigma}{\delta\Theta_{\mu}} & = & \partial_{\mu}b\;.\label{seqss}
\end{eqnarray}
Notice that all terms in the right hand side of eqs.~(\ref{seqss}) are linear breakings, which will not be affected by quantum corrections \cite{Piguet:1995er}. As a consequence, these equations  imply that the most general local invariant counterterm does not depend on $(\eta,\Upsilon,\Theta )$.

As shown in  \cite{Capri:2020ppe}, besides the previous Ward identities, the complete action $\Sigma$ enjoys two additional powerful Ward identities
which have far reaching consequences at the quantum level, namely
\begin{eqnarray}
\int d^{4}x\left(\frac{\delta\Sigma}{\delta h}-\lambda v\frac{\delta\Sigma}{\delta J}\right)-\frac{\partial\Sigma}{\partial v} & = & \int d^{4}xv\left(J-2\eta\right)\;, \label{hequation}
\end{eqnarray}
and
\begin{equation}
\frac{\delta \Sigma}{\delta A_\mu} - 2 e \frac{\delta \Sigma}{\delta \Omega_\mu} - e \Omega_\mu \frac{\delta \Sigma}{\delta J} = -\partial_\nu F_{\nu\mu} - i \partial_\mu b + \frac{e v^2}{2} \Omega_\mu + \partial^2 \Upsilon_\mu - \partial_\mu \partial_\nu \Upsilon_{\nu} \;. \label{Widom}
\end{equation}
It is worth emphasizing that, unlike the Ward identities (\ref{slavnov})-(\ref{seqss}), which can be written down  independently from the introduction of the composite operators $(O(x),V_{\mu}(x))$, the additional Ward identities (\ref{hequation}) and (\ref{Widom}) can be obtained only when the composite operators are introduced in the action from the beginning. As such, these two Ward identities will express non-trivial features of the correlation functions of the two composite operators $(O(x),V_{\mu}(x))$, as it will be illustrated later on.

 The Ward identity (\ref{hequation}) will lead to the already mentioned relation \eqref{Wennw} (see later in \eqref{wev}), explicitly connecting the vacuum energy ${\cal E}_{v}$ to the tadpole diagrams in a very simple and apparent way. This is one of the main novelties of our paper. On the other hand, as underlined in  \cite{Capri:2020ppe}, the second Ward identity (\ref{Widom}) expresses the fact that the gauge invariant vector operator $V_\mu(x)$ is a conserved current, a property which will be addressed in detail in the next section. Finally, let us observe that also both eqs.~(\ref{hequation}) and (\ref{Widom})
display a harmless linear breaking.

\section{Investigating the nature of the gauge invariant vector operator $V_\mu(x)$} \label{S3}
Before addressing the issue of the vacuum energy ${\cal E}_{v}$, let us investigate the nature of the gauge invariant vector operator $V_\mu$. As stated in the previous section, see \cite{Capri:2020ppe}, the composite operators $V_\mu(x)$ has the meaning of a conserved current. For a better understanding of this feature, we  rewrite the global Ward identity (\ref{gboW})  as
\begin{equation}
\int d^4 x\; {\cal W}(x) \Sigma = 0\;, \label{Wg}
\end{equation}
where ${\cal W}(x)$ stands for the local operator
\begin{eqnarray}
{\cal W}(x) = \left[-\rho(x)\frac{\delta}{\delta h(x)}+\left(v+h(x)\right)\frac{\delta}{\delta\rho(x)}-R(x)\frac{\delta}{\delta L(x)}+L(x)\frac{\delta}{\delta R(x)}\right]  \;. \label{Wop}
\end{eqnarray}
Therefore, from the Noether theorem, we get
\begin{equation}
{\cal W}(x) \Sigma = \partial_{\mu} J_\mu(x) \;,
\end{equation}
where, after a quick algebraic calculation, the current $J_\mu$ reads
\begin{equation}
J_\mu(x) = - \left( \Omega_\mu(x) O(x) + \frac{v^2}{2} \Omega_\mu(x) + 2 V_\mu(x) \right) \;. \label{currj}
\end{equation}
Setting thus all external sources $(J,\eta,\Omega_{\mu},\Upsilon_{\mu},\Theta_{\mu} )$ to zero , one gets
\begin{equation}
\left[-\rho(x)\frac{\delta}{\delta h(x)}+\left(v+h(x)\right)\frac{\delta}{\delta\rho(x)} \right] (S_{\rm Higgs}+S_{\rm gf}) = - 2 \partial_\mu V_\mu(x) \;, \label{Vcur}
\end{equation}
which shows that the gauge invariant vector operator $V_\mu(x)$ is nothing but the conserved Noether current corresponding to the global $U(1)$ invariance, manifestly preserved in the Landau gauge.

Furthermore, having introduced the two composite operators $(O(x), V_\mu(x)$ in the starting action $\Sigma$ from the beginning, equation (\ref{Vcur}) can be directly converted into a local Ward identity, namely
\begin{equation}
\left[-\rho(x)\frac{\delta\Sigma}{\delta h(x)}+\left(v+h(x)\right)\frac{\delta\Sigma}{\delta\rho(x)}-R(x)\frac{\delta\Sigma}{\delta L(x)}+L(x)\frac{\delta\Sigma}{\delta R(x)}\right] = -\partial_\mu \left(\Omega_\mu (x) \frac{\delta \Sigma}{\delta J(x)} + 2 \frac{\delta \Sigma}{\delta \Omega_\mu (x)} +
 \frac{v^2}{2} \Omega_\mu(x) \right)  \;. \label{W1}
\end{equation}
Finally, let us also observe that, upon making use of (\ref{Widom}), equation (\ref{W1}) becomes the familiar linearly broken local $U(1)$  gauge Ward identity
\begin{equation}
\partial_\mu  \frac{\delta \Sigma}{\delta A_\mu (x)} + e \left[-\rho(x)\frac{\delta \Sigma}{\delta h(x)}+\left(v+h(x)\right)\frac{\delta \Sigma}{\delta\rho(x)}-R(x)\frac{\delta \Sigma}{\delta L(x)}+L(x)\frac{\delta \Sigma}{\delta R(x)} \right] = -i \partial^2 b (x) \;, \label{locWg}
\end{equation}
which follows by anti-commuting the Slavnov-Taylor Ward identity  (\ref{slavnov}) with the ghost Ward identities (\ref{eq:cequation}).

\section{Revisiting the most general form of the invariant counterterm: adding the vacuum energy
counterterm}\label{VC}

In this section we revise the construction of the more general invariant counterterm compatible with the whole set of Ward indentities. We shall pay particular attention to the inclusion\footnote{In the previous work \cite{Capri:2020ppe},  the vacuum counterterm $v^4$ was not needed, as it was not  entering the correlation functions $\langle O(p) h(-p) \rangle$  and $\langle V_\mu(p) A_\nu(-p) \rangle$ whose study was the main goal of the work.} of the vacuum counterterm of the type $ v^4$,  which is generated from one loop onwards and which is needed to  properly renormalize the vacuum energy. Usually, such a kind of counterterm is not introduced in the analysis of the counterterm, as it is not captured nor constrained by the Ward identities which are expressed in terms of functional derivates with respect to the fields and external sources. Nevertheless, due to the introduction of the operators $(O(x),V_\mu(x))$ in the starting action, it has been possible to derive the two additional Ward identities (\ref{hequation}) and  (\ref{Widom}). In particular, we point out the appearance of the term $\frac{\partial \Sigma}{\partial v}$ which, as we shall see,
will capture the dependence of the theory from  the vacuum counterterm $ v^4$, relating it to the tadpole diagrams at the quantum level in a BRST  invariant fashion, see  \cite{Becchi:1974md,Becchi:1974xu,Kraus:1995jk,Haussling:1996rq}. In other words, the Ward identity (\ref{hequation}) will ensure in an explicit way that the vanishing condition of the tadpole diagrams is related to the minimization procedure of the vacuum energy. A whole subsection will be devoted to this issue. We highlight again that such a possibility depends crucially from the a priori introduction of the two gauge invariant composite operators $(O(x),V_\mu(x))$ in the starting action.

In order to characterize the most general local invariant counterterm, we follow the algebraic renormalization setup \cite{Piguet:1995er} and perturb the starting action $\Sigma$, i.e.
$\Sigma \rightarrow (\Sigma + \epsilon \Sigma^{\rm ct})$ with $\epsilon$ being an expansion parameter. In agreement with the power counting,  $\Sigma^{\rm ct}$ is
an integrated local polynomial in the fields and  external sources with dimension four, invariant under charge conjugation and having vanishing ghost number. Demanding then  that  the perturbed action, $(\Sigma + \epsilon \Sigma^{\rm ct})$, fulfills to the first order in the expansion parameter $\epsilon$  the same Ward identities of the action $\Sigma$, eqs.~(\ref{slavnov})- (\ref{Widom}), one gets the following conditions
\begin{eqnarray}
\frac{\delta\Sigma^{\rm ct}}{\delta b} & = & \frac{\delta\Sigma^{\rm ct}}{\delta\overline{c}}=\frac{\delta\Sigma^{ct}}{\delta c}=0 \;, \label{cd1}
\end{eqnarray}
as well as
\begin{eqnarray}
\frac{\delta\Sigma^{\rm ct}}{\delta\eta} = \frac{\delta\Sigma^{\rm ct}}{\delta\Theta_{\mu}}=\frac{\delta\Sigma^{ct}}{\delta\Upsilon_{\mu}}=0 \;.
\label{cd2}
\end{eqnarray}
Since $\Sigma^{\rm ct}$ is independent from the ghosts $(c,\overline{c})$, it immediately follows that, due to the fact that the sources $(L,R)$ have ghost number $-1$, they cannot give rise to a dimension four quantity with vanishing ghost number, namely
\begin{eqnarray}
\frac{\delta\Sigma^{\rm ct}}{\delta L} = \frac{\delta\Sigma^{\rm ct}}{\delta R}=0\;.\label{eq:sourcesequation}
\end{eqnarray}
Therefore
\begin{equation}
\Sigma^{\rm ct} = \Sigma^{ct}(A,h,\rho,v,J,\Omega) \;. \label{ct1}
\end{equation}
The result (\ref{eq:sourcesequation}) simplifies very much the Slavnov-Taylor
identity, which takes the simpler form
\begin{eqnarray}
s\Sigma^{\rm ct} = 0\;. \label{stct}
\end{eqnarray}
From equations (\ref{hequation}) and (\ref{Widom}) one obtains  two additional conditions
\begin{eqnarray}
\frac{\delta\Sigma^{ct}}{\delta A_{\mu}}-2e\frac{\delta\Sigma^{ct}}{\delta\Omega_{\mu}}-e\Omega_{\mu}\frac{\delta\Sigma^{ct}}{\delta J} = 0 \;,
\label{aequationct}
\end{eqnarray}
and
\begin{eqnarray}
\int d^{4}x\left(\frac{\delta\Sigma^{ct}}{\delta h}-\lambda v\frac{\delta\Sigma^{ct}}{\delta J}\right)-\frac{\partial\Sigma^{ct}}{\partial v} = 0\,.
\label{hequationct}
\end{eqnarray}
After some algebraic calculation, from equations (\ref{cd1})-(\ref{stct}) one gets,
\begin{eqnarray}
\Sigma^{ct}  = & &  \int  d^{4}x  \left\{ a_{0}\frac{1}{4}F_{\mu\nu}F_{\mu\nu}+a_{1}\left(D_{\mu}\varphi\right)^{*}D_{\mu}\varphi+a_{2}\frac{\lambda}{2}\left(\varphi^{*}\varphi-\frac{v^{2}}{2}\right)^{2}+(\delta a)\frac{v^{4}}{4}\right. \nonumber \\
 &  &\qquad  +\;\delta\sigma\frac{v^{2}}{2}\left(h^{2}+2vh+\rho^{2}\right)
  +b_{1}JO+b_{2}Jv^{2}+c_{1}\Omega_{\mu}V_{\mu}+c_{2}\Omega_{\mu}\partial_{\nu}F_{\nu\mu} \nonumber \\
 &  & \qquad +\;d_{1}\Omega_{\mu}\Omega_{\mu}\Omega_{\nu}\Omega_{\nu}+d_{2}\Omega_{\mu}\partial^{2}\Omega_{\mu}+d_{3}\Omega_{\mu}\partial_{\mu}\partial_{\nu}\Omega_{\nu} \nonumber \\
 &  & \qquad  \left.+\;d_{4}v^{2}\Omega_{\mu}\Omega_{\mu}+d_{5}O\Omega_{\mu}\Omega_{\mu}+d_{6}J^{2}+d_{7}J\Omega_{\mu}\Omega_{\mu}\right\}  \;, \label{ctv1}
\end{eqnarray}
where one notices the introduction of the vacuum counterterm $(\delta a) v^{4}$.

Imposing now the two conditions (\ref{aequationct}) and (\ref{hequationct}), we have
\begin{eqnarray}
\Sigma^{ct} & = & \int d^{4}x\left\{ a_{0}\left(\frac{1}{4}F_{\mu\nu}F_{\mu\nu}-\frac{1}{2e}\Omega_{\mu}\partial_{\nu}F_{\nu\mu}-\frac{1}{8e^{2}}\Omega_{\mu}\partial^{2}\Omega_{\mu}+\frac{1}{8e^{2}}\Omega_{\mu}\partial_{\mu}\partial_{\nu}\Omega_{\nu}\right)\right. \nonumber \\
 &  & +a_{1}\left(\left(D_{\mu}\varphi\right)^{*}D_{\mu}\varphi+\Omega_{\mu}V_{\mu}+\frac{1}{8}v^{2}\Omega_{\mu}\Omega_{\mu}+\frac{1}{4}O\Omega_{\mu}\Omega_{\mu}\right) \nonumber \\
 &  & +a_{2}\left[\frac{\lambda}{2}\left(\varphi^{*}\varphi-\frac{v^{2}}{2}\right)^{2}+JO-\frac{1}{4}O\Omega_{\mu}\Omega_{\mu}+\frac{1}{32\lambda}\left(\Omega_{\mu}\Omega_{\mu}\Omega_{\nu}\Omega_{\nu}+16J^{2}-8J\Omega_{\mu}\Omega_{\mu}\right)\right] \nonumber \\
 &  & +(\delta a)\left[\frac{v^{4}}{4}+\frac{1}{16\lambda^{2}}\left(\Omega_{\mu}\Omega_{\mu}\Omega_{\nu}\Omega_{\nu}+16J^{2}-8J\Omega_{\mu}\Omega_{\mu}\right)-\frac{1}{\lambda}\left(Jv^{2}-\frac{1}{4}v^{2}\Omega_{\mu}\Omega_{\mu}\right)\right] \nonumber \\
 &  & +\delta\sigma\left[\frac{v^{2}}{2}\left(h^{2}+2vh+\rho^{2}\right)+\frac{1}{\lambda}\left(Jv^{2}-\frac{1}{4}v^{2}\Omega_{\mu}\Omega_{\mu}-2JO+\frac{1}{2}O\Omega_{\mu}\Omega_{\mu}\right)\right. \nonumber \\
 &  & \left.\left.-\frac{1}{8\lambda^{2}}\left(\Omega_{\mu}\Omega_{\mu}\Omega_{\nu}\Omega_{\nu}+16J^{2}-8J\Omega_{\mu}\Omega_{\mu}\right)\right]\right\} \nonumber  \\;. \label{fctt}
\end{eqnarray}
with $(a_0, a_1, a_2,\delta \sigma, \delta a)$ free parameters.

As already underlined in \cite{Capri:2020ppe}, the counterterm (\ref{fctt}) exhibits  a few properties worth underlining. The first one is the presence of the term $\Omega_{\mu}\partial_{\nu}F_{\mu\nu}$, with $\Omega_\mu$ being the source coupled to the vector operator $V_\mu$. As shown in \cite{Capri:2020ppe},  this term gives rise to the mixing between the operators $V_\mu$ and $\partial_{\nu}F_{\mu\nu}$. The second feature concerns the well known presence \cite{Becchi:1974md,Becchi:1974xu,Kraus:1995jk,Haussling:1996rq} of the BRST invariant counterterm $(\delta\sigma)\frac{v^{2}}{2} (h^2 +2vh +\rho^2)$. As we shall see explicitly in  the next section, both  coefficients $(\delta\sigma)$ and $\delta a $ can be fixed, order by order in the $\hbar $ expansion, so as to ensure the minimization procedure for the vacuum energy ${\cal E}_{v}$ and the requirement of vanishing tadpoles. Finally, let us notice  that the Ward identities allow for the presence of higher order terms in the external sources, like ($\Omega_{\mu}\partial^{2} \Omega_{\mu}$, $\Omega^4$, $J\Omega^2$, $\ldots$). These terms, which originate from one loop onwards, are needed to properly renormalize the two-point correlation function of the composite operators as, for example, $\langle V_\mu(p) V_\nu(-p) \rangle$,  whose explicit one-loop computation will be presented in details.

After having obtained the most general form of the local invariant BRST counterterm, eq.~(\ref{fctt}), we can obtain the  bare action. For such a goal we  look at how the counterterm (\ref{fctt}) can be reabsorbed into the tree level action $\Sigma$, namely
\begin{eqnarray}
 \Sigma+\epsilon \Sigma^{\rm ct}= \Sigma_{\rm bare} + O(\epsilon^2) \;, \label{bare}
\end{eqnarray}
 where
\begin{eqnarray}
 \Sigma_{\rm bare} & = & \Sigma\left(A_{0\mu},h_{0},\rho_{0},b_{0},c_{0},\overline{c}_{0},v_{0},e_{0},\lambda_{0},J_{0},\eta_{0},\Omega_{0\mu},\Upsilon_{0\mu},\Theta_{0\mu},L_{0},R_{0}\right) \nonumber \\
 &  & +\int d^{4}x\delta\sigma_{0}\frac{v_{0}^{2} }{2}\left(h_{0}^{2}+2v_{0}h_{0}+\rho_{0}^{2}\right) \nonumber \\
 &  & +\int d^{4}x\; \left(Z_{A}-1\right)\left(-\frac{1}{8e_{0}^{2}}\Omega_{0\mu}\partial^{2}\Omega_{0\mu}+\frac{1}{8e_{0}^{2}}\Omega_{0\mu}\partial_{\mu}\partial_{\nu}\Omega_{0\nu}\right) \nonumber \\
 &  & +\int d^4x\; \left(Z_{h}-1\right)\left(\frac{1}{8}v_{0}^{2}\Omega_{0\mu}\Omega_{0\mu}+\frac{1}{4}O_{0}\Omega_{0\mu}\Omega_{0\mu}\right)\nonumber \\
 &  & + \int^44 x\; \left(Z_{\lambda}+2Z_{h}-3\right)\left[-\frac{1}{4}O_{0}\Omega_{0\mu}\Omega_{0\mu}+\frac{1}{32\lambda_{0}}\left(\Omega_{0\mu}\Omega_{0\mu}\Omega_{0\nu}\Omega_{0\nu}+16J_{0}^{2}-8J_{0}\Omega_{0\mu}\Omega_{0\mu}\right)\right] \nonumber \\
 &  & + \int d^4x\; \delta\sigma_{0}\left[\frac{1}{\lambda_{0}}\left(-\frac{1}{4}v_{0}^{2}\Omega_{0\mu}\Omega_{0\mu}+\frac{1}{2}O_{0}\Omega_{0\mu}\Omega_{0\mu}\right) -\frac{1}{8\lambda_{0}^{2}}\left(\Omega_{0\mu}\Omega_{0\mu}\Omega_{0\nu}\Omega_{0\nu}+16J_{0}^{2}-8J_{0}\Omega_{0\mu}\Omega_{0\mu}\right)\right] \nonumber \\
  & & + \int d^4x\; {(\delta a)}_0\left[\frac{v^{4}_0}{4}+\frac{1}{16\lambda^{2}_0}\left(\Omega_{0\mu}\Omega_{0\mu}\Omega_{0\nu}\Omega_{0\nu}+
  16J^{2}_0-8J_0 \Omega_{0\mu}\Omega_{0\mu}\right)+ \frac{v^{2}_0}{4\lambda_0} \Omega_{0\mu}\Omega_{0\mu}\right] \nonumber \\  \label{bact}
\end{eqnarray}
with
\begin{eqnarray}
A_{0\mu} & = & Z_{A}^{\frac{1}{2}}A_{\mu}\label{eq:renA} \;, \nonumber \\
h_{0} & = & Z_{h}^{\frac{1}{2}}h    \;, \nonumber \\
\rho_{0} & = & Z_{\rho}^{\frac{1}{2}}\rho \;, \nonumber \\
v_0 & = & Z_{v}^{\frac{1}{2}} v \;, \nonumber \\
b_{0} & = & Z_{b}^{\frac{1}{2}}b \;, \nonumber \\
c_{0} & = & Z_{c}^{\frac{1}{2}}c  \;, \nonumber \\
\overline{c}_{0} & = & Z_{\overline{c}}^{\frac{1}{2}}  \overline{c} \;, \nonumber \\
e_{0} & = & Z_{e}e \;, \nonumber \\
\lambda_{0} & = & Z_{\lambda}\lambda \;, \nonumber \\
L_{0} & = & Z_{L}L \;, \nonumber \\
R_{0} & = & Z_{R}R \;, \nonumber \\
\Theta_{\mu0} & = & Z_{\Theta}   \Theta \;, \label{z1}
\end{eqnarray}
and
\begin{eqnarray}
\left(\begin{array}{c}
\Omega_{0\mu}\\
\Upsilon_{0\mu}
\end{array}\right) & = & \left(\begin{array}{cc}
Z_{\Omega\Omega} & Z_{\Omega\Upsilon}\\
Z_{\Upsilon\Omega} & Z_{\Upsilon\Upsilon}
\end{array}\right)\left(\begin{array}{c}
\Omega_{\mu}\\
\Upsilon_{\mu}
\end{array}\right) \; \nonumber \\
\left(\begin{array}{c}
J_{0}\\
\eta_{0}
\end{array}\right) & = & \left(\begin{array}{cc}
Z_{JJ} & Z_{J\eta}\\
Z_{\eta J} & Z_{\eta\eta}
\end{array}\right)\left(\begin{array}{c}
J\\
\eta
\end{array}\right)\label{eq:renJ} \;.  \label{matrix}
\end{eqnarray}
In particular, the matrix form of equations (\ref{matrix}) expresses, in terms of the corresponding external sources
$(\Omega_\mu, \Upsilon_\mu, J, \eta)$,  the mixing between the quantities $V_\mu$ and $(\partial_\nu F_{\nu \mu})$ as well as between $O(x)$ and $v^2$.

A direct inspection of equation (\ref{bare}) yields
\begin{eqnarray}
Z_{A}^{\frac{1}{2}} & = & Z_{e}^{-1}=1+\frac{1}{2}\epsilon  a_{0} \;, \nonumber \\
Z_{h}^{\frac{1}{2}} & = & Z_{\rho}^{\frac{1}{2}}=Z_{v}^{\frac{1}{2}}=1+\frac{1}{2}\epsilon a_{1} \;, \nonumber \\
Z_{\lambda} & = & 1+\epsilon \left(a_{2}-2a_{1}\right) \;, \nonumber \\
Z_{c}^{\frac{1}{2}} & = & Z_{\overline{c}}^{-\frac{1}{2}} \;, \nonumber \\
Z_{\Theta} & = & Z_{b}^{-\frac{1}{2}}=Z_{A}^{\frac{1}{2}}\;, \nonumber \\
Z_{L} & = & Z_{R}=Z_{e}^{-1}Z_{h}^{-\frac{1}{2}}Z_{c}^{-\frac{1}{2}} \;, \nonumber \\
Z_{\Omega\Omega} & = & 1\;, \nonumber \\
Z_{\Omega\Upsilon} & = & 0\;, \nonumber \\
Z_{\Upsilon\Omega} & = & -\frac{1}{2e}\epsilon  a_{0}= -\frac{1}{2e} \left(Z_{A}-1\right)\;, \nonumber \\
Z_{\Upsilon\Upsilon} & = & Z_{A}^{-\frac{1}{2}}=1-\frac{1}{2}\epsilon a_{0} \;, \nonumber \\
Z_{JJ} & = & 1+\epsilon \left(a_{2}-a_{1}-2\frac{\delta \sigma}{\lambda}\right) \;, \nonumber \\
Z_{J\eta} & = & 0 \;, \nonumber \\
Z_{\eta J} & = & \epsilon \left( \frac{\delta \sigma}{\lambda}-\frac{\delta a}{\lambda}\right) \;, \nonumber \\
Z_{\eta\eta} & = & Z_{h}^{-1}=1-\epsilon  a_{1} \;, \label{zzz1}
\end{eqnarray}
and
\begin{equation}
(\delta \sigma)_0 = \epsilon (\delta \sigma) \;. \qquad {\delta a}_0 = \epsilon {\delta a } \label{ds}
\end{equation}
Thus, for the bare action, we get\footnote{Since we are not interested in the calculation of Green's functions with insertions of the BRST exact operators $(sh, s\rho)$, from now on we shall set to zero  the corresponding external sources, i.e.
$L=R=0$.}
\begin{eqnarray}
\Sigma_{bare} & = & \int d^{4}x\left( \frac{1}{4}F_{0\mu\nu}F_{0\mu\nu}+\left(D_{0\mu}\varphi_{0}\right)^{\ast}\left(D_{0\mu}\varphi_{0}\right)+\frac{\lambda_{0}}{2}\left(\varphi_{0}^{\ast}\varphi_{0}-\frac{v_{0}^{2}}{2}\right)^{2} \right) \nonumber \\
 &  & + \int d^4x\; \left( \overline{c}_{0}\partial^{2}c_{0}+ib_{0}\partial_{\mu}A_{0\mu} +J_{0}O_{0}+\eta_{0}v_{0}^{2}
  +\Omega_{0\mu}V_{0\mu}+\Upsilon_{0\mu}\partial_{\nu}F_{0\nu\mu} \right) \nonumber \\
 &  & +  \int d^4x \left( \frac{(\delta \sigma)_0}{2} v^{2}_0\left(h^{2}_0+2v_0h_0+\rho^{2}_0\right) \right) \; \nonumber \\
 & &+\int d^{4}x \left(Z_{A}-1\right)\left(-\frac{1}{8e_{0}^{2}}\Omega_{0\mu}\partial^{2}\Omega_{0\mu}+\frac{1}{8e_{0}^{2}}\Omega_{0\mu}\partial_{\mu}\partial_{\nu}\Omega_{0\nu}\right) \nonumber \\
 &  & +\int d^4x\; \left(Z_{h}-1\right)\left(\frac{1}{8}v_{0}^{2}\Omega_{0\mu}\Omega_{0\mu}+\frac{1}{4}O_{0}\Omega_{0\mu}\Omega_{0\mu}\right)\nonumber \\
 &  & +\int d^4 x\; \left(Z_{\lambda}+2Z_{h}-3\right)\left[-\frac{1}{4}O_{0}\Omega_{0\mu}\Omega_{0\mu}+\frac{1}{32\lambda_{0}}\left(\Omega_{0\mu}\Omega_{0\mu}\Omega_{0\nu}\Omega_{0\nu}+16J_{0}^{2}-8J_{0}\Omega_{0\mu}\Omega_{0\mu}\right)\right] \nonumber \\
 &  & +\int d^4x\; \delta\sigma_{0}\left[\frac{1}{\lambda_{0}}\left(  -\frac{1}{4}v_{0}^{2}\Omega_{0\mu}\Omega_{0\mu}+\frac{1}{2}O_{0}\Omega_{0\mu}\Omega_{0\mu}\right) \right] \nonumber \\
 & &+\int d^4x\; \delta\sigma_{0}\left[  -\frac{1}{8\lambda_{0}^{2}}\left(\Omega_{0\mu}\Omega_{0\mu}\Omega_{0\nu}\Omega_{0\nu}+16J_{0}^{2}-8J_{0}\Omega_{0\mu}\Omega_{0\mu}\right)\right] \nonumber \\
 \nonumber \\
  & & + \int d^4x\; {(\delta a)}_0\left[\frac{v^{4}_0}{4}+\frac{1}{16\lambda^{2}_0}\left(\Omega_{0\mu}\Omega_{0\mu}\Omega_{0\nu}\Omega_{0\nu}+
  16J^{2}_0-8J_0 \Omega_{0\mu}\Omega_{0\mu}\right) +\frac{v^{2}_0}{4 \lambda_0}  \Omega_{0\mu}\Omega_{0\mu}  \right] \nonumber \\   , \label{baction}
\end{eqnarray}
with
\begin{equation}
\varphi_0  =  \frac{Z_h^{1/2}}{\sqrt{2}}\left(v+h+i\rho\right)\;. \label{phi0}
\end{equation}
Let us end this section by giving, for completeness,  the explicit values of the $Z$ factors \cite{Capri:2020ppe}, as evaluated at one-loop order in the $\overline{\text{MS}}$ scheme, working  with dimensional regularization where $d=(4 - \varepsilon)$. Let us begin with the $Z$ factors of $(h, \rho,v)$. We have
\begin{eqnarray}
Z_{h}  =  Z_{\rho}=Z_{v}  \;, \label{eql}
\end{eqnarray}
where, at the one-loop order
\begin{eqnarray}
Z_{h}^{\left(1\right)} & = & \frac{3e^{2}}{16\pi^{2}}\left(\frac{2}{\varepsilon}-\gamma+\ln\left(4\pi\right)\right) \;. \label{zhh}
\end{eqnarray}
The equality of the $Z$ factors for $(h, \rho,v)$ follows from the global invariance, eq.~(\ref{Wg}), which is manifestly preserved in the Landau gauge. For $Z_{\lambda}^{\left(1\right)}$ we get
\begin{eqnarray}
Z_{\lambda}^{\left(1\right)} & = & \frac{1}{16\pi^{2}}\left(5\lambda+6\frac{e^{4}}{\lambda}-6e^{2}\right)\left(\frac{2}{\varepsilon}-\gamma+\ln\left(4\pi\right)\right) \;, \label{zllt}
\end{eqnarray}
while  $ Z_e$ and $ Z_A^{1/2} $  turn out to be
\begin{equation}
  Z_e Z_A^{1/2} = 1 \;, \label{eaa}
\end{equation}
with
\begin{equation}
Z_{A}^{\left(1\right)}  =  -\frac{e^{2}}{48\pi^{2}}\left(\frac{2}{\varepsilon}-\gamma+\ln\left(4\pi\right)\right) \;, \label{zaat1}
\end{equation}
expressing a  well known feature of the Abelian U(1) models, see, for example,  \cite{Itzykson:1980rh}. Finally,  the $Z$ factors off the sources coupled to the composite operators $(O(x), V_\mu(x))$ and to $\partial_\nu F_{\mu\nu}$ are given by
\begin{eqnarray}
Z_{JJ}^{\left(1\right)} & = & \frac{1}{16\pi^{2}}\left(2\lambda-3e^{2}\right)\left(\frac{2}{\varepsilon}-\gamma+\ln\left(4\pi\right)\right)\;. \label{zjj}
\end{eqnarray}
and
\begin{eqnarray}
Z_{\Omega\Omega}^{\left(1\right)} & = & 0  \;, \label{rs1}\\
Z_{\Upsilon\Omega}^{\left(1\right)} & = & \frac{e}{96\pi^{2}}\left(\frac{2}{\varepsilon}-\gamma+\ln\left(4\pi\right)\right) \;. \label{rs2}
\end{eqnarray}
Notice that  eq.~(\ref{rs1}) is in perfect agreement with the general output of the Ward identities, eqs.~(\ref{zzz1}), namely $Z_{\Omega\Omega}=1$, whose origin is due to the observation that  $V_\mu$ is a conserved current of a global symmetry. It is standard textbook material that conserved currents do not need renormalization, expressed by $Z_{\Omega\Omega}=1$ \cite{Collins:1984xc}. However, the situation is bit more subtle than that, apart from the potential mixing with other operators. Indeed, the algebraic analysis also makes obvious the need for contact counterterms (corresponding to the pure source terms) which are necessary to render finite correlation functions involving currents. These contact terms are in particular necessary for the polynomial subtractions when passing to a convergent K\"all\'{e}n-Lehmann spectral representation of the two-point correlation function, see \cite{weinberg}.

\section{Tadpoles, vacuum energy and the condensate $\langle O \rangle$}\label{OC}
This section is devoted to a detailed analysis of the BRST invariant counterterms $ \left( \frac{(\delta \sigma)_0}{2} v^{2}_0\left(h^{2}_0+2v_0h_0+\rho^{2}_0\right) \right) $ and $(\delta a)_0 \frac{v^{4}_0}{4}$, which appear in the most general expression for the bare action, eq.~(\ref{baction}), in the light of the exact constraint \eqref{Wennw} which follows from the Ward identity \eqref{hequation}.

Let us start first by discussing the counterterm stemming from $ \left( \frac{(\delta \sigma)_0}{2} v^{2}_0\left(h^{2}_0+2v_0h_0+\rho^{2}_0\right) \right) $, which presence is a well established feature of the Higgs model \cite{Becchi:1974md,Becchi:1974xu,Kraus:1995jk,Haussling:1996rq} . Following \cite{Becchi:1974md,Becchi:1974xu,Kraus:1995jk,Haussling:1996rq}, this counterterm is fixed, order by order in the loop expansion, by requiring the vanishing of the tadpoles, namely by imposing
\begin{equation}
\langle h\rangle = 0\;. \label{rad}
\end{equation}
For example, at one-loop, summing up all non-vanishing contributions, see Figure \ref{1looptadpolea}, one gets, again using the $\overline{\text{MS}}$ scheme in $d=(4 - \varepsilon)$ dimensions,
\begin{equation}
\left\langle h\left(x\right)\right\rangle_{\rm 1-loop} = \frac{3 \lambda v}{2} \chi\left(m^{2}_h\right) + e^2 v (d-1)\chi\left(m^{2}\right) + (\delta\sigma)^{\left(1\right)} v^3 = 0 \;, \label{vtdp}
\end{equation}
where $\chi\left(M^{2}\right)$ stands for
\begin{eqnarray}
\chi\left(M^{2}\right)  =  \int\frac{d^{d}k}{\left(2\pi\right)^{d}}\frac{1}{k^{2}+M^{2}}
  =  \frac{1}{\left(4\pi\right)^{\frac{d}{2}}}\Gamma\left(1-\frac{d}{2}\right)\left(M^{2}\right)^{\frac{d}{2}-1} \;. \label{chim}
\end{eqnarray}
Therefore,  $(\delta\sigma)^{\left(1\right)} $ becomes
\begin{eqnarray}
(\delta\sigma)^{\left(1\right)}  =  \frac{1}{v^{2}}\left(-e^{2}\left(d-1\right)\chi\left(m^{2}\right)-\frac{3}{2}\lambda\chi\left(m_{h}^{2}
\right)\right) \;. \label{tdp2}
\end{eqnarray}
In particular, employing the subtraction procedure of the $\overline{\text{MS}}$ scheme, for the divergent part, one has:
\begin{eqnarray}
 (\delta\sigma)^{\left(1\right)}_{\rm div} =  \frac{1}{\left(4\pi\right)^{2}}\frac{1}{v^{2}}\left(3e^{2}m^{2}+\frac{3}{2}\lambda m_{h}^{2}\right)
\left(\frac{2}{\varepsilon}-\gamma+\ln\left(4\pi\right)\right) \;. \label{tdp3}
\end{eqnarray}

\begin{figure}[h]
\begin{center}
\includegraphics[width=.5\textwidth,angle=0]{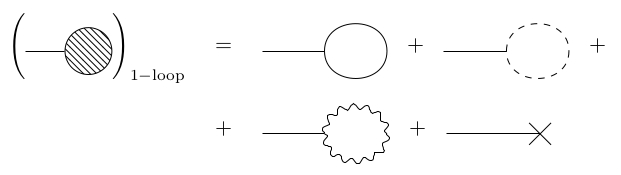}
\caption{One-loop diagrams contributing to the one-point Green's function $\langle h \rangle$  of the Higgs field $h$. Curly lines refer to the gauge field, solid ones to the Higgs field and dashed ones to the Goldstone field.
}
\label{1looptadpolea}
\end{center}
\end{figure}
\noindent Let us turn now to the vacuum energy ${\cal E}_{v}$. At one-loop order\footnote{The classical vacuum energy is zero as $\varphi=\frac{v}{\sqrt 2}$ corresponds to the classical minimum.}, one easily gets from the bubble graph
\begin{equation}
{\cal E}_{v} = \frac{1}{2} \int d^dk \log(k^2+ m^2_h) +  \frac{d-1}{2} \int d^dk \log(k^2+ m^2)  +{(\delta a)}^1 \;\frac{ v^4}{4} \;. \label{ven}
\end{equation}
Taking the derivative of eq.~(\ref{ven}) with respect to $v$, it follows
\begin{equation}
\frac{\partial {\cal E}_{v} }{\partial v} = \lambda v \chi\left(m_h^{2}\right) + e^2 v (d-1)  \chi\left(m^{2}\right) + {(\delta a)}^1 v^3 \;, \label{dev}
\end{equation}
which, using eq.~(\ref{tdp2}), can be rewritten as
\begin{equation}
\frac{\partial {\cal E}_{v} }{\partial v} = \langle h \rangle_{\rm 1-loop}  + {(\delta a)}^1 v^3 - (\delta\sigma)^{\left(1\right)}v^3 - \frac{\lambda v}{2} \chi\left(m_h^{2}\right) \;. \label{rdev}
\end{equation}
Therefore,  we determine the counterterm $ {(\delta a)}^1$ in such a way that
\begin{equation}
{(\delta a)}^1  = (\delta\sigma)^{\left(1\right)} + \frac{\lambda }{2v^2} \chi\left(m_h^{2}\right)  = \frac{1}{v^2} \left( - e^2 (d-1)  \chi\left(m^{2}\right) -\lambda  \chi\left(m_h^{2}\right) \right) \;. \label{da1}
\end{equation}
As a consequence, we have
\begin{equation}
\frac{\partial {\cal E}_{v} }{\partial v} = \left\langle h\right\rangle_{\rm 1-loop} = 0  \;, \label{rdev1}
\end{equation}
expressing the important fact that the vacuum energy ${\cal E}_{v}$ is minimized at $v$.

 To be more precise, one can fix the two free BRST invariant counterterms, $(\delta \sigma)_0$, $(\delta a)_0$, order by order in the loop expansion, in such a way that the two conditions, namely: vanishing of the tadpoles and  minimization condition for the vacuum energy ${\cal E}_{v}$, are simultaneously fulfilled.

It is worth emphasizing that characterizing the counterterm ${(\delta a)}^1 $ in such a way that equation (\ref{rdev1}) holds has a deep relationship with the Ward identity (\ref{hequation}). In fact, by direct inspection of the bare action, eq.~(\ref{bact}), one realizes that the perturbative  condensate $\langle O\rangle_{pert}$ will vanish if the two BRST invariant counterterms $(\delta \sigma)_0$, $(\delta a)_0$ are fixed as described. Indeed, we get
\begin{equation}
\langle O \rangle_{pert} = \frac{1}{2} \langle h^2(x) + 2vh(x) + \rho^2(x) \rangle_{pert} = 0 \;. \label{vancond}
\end{equation}
This follows by noticing that $\langle O\rangle_{pert}$ is given by
\begin{equation}
\langle O\rangle_{pert} = \frac{\delta Z^c}{\delta J}\Big|_{\rm sources =0} \;, \label{OZ}
\end{equation}
where $Z^c$  denotes the generator of the connected Green functions. A simple way to evaluate $\langle O\rangle_{pert}$ is that of computing the Feynman diagrams which have as unique external leg the source $J$. After that, one differentiates with respect to $J$ in agreement with eq.~(\ref{OZ}), thus obtaining $\langle O\rangle_{pert}$.

A quick look at the bare action, eq.~(\ref{baction}), reveals that the terms linear in the source $J$ are hiding in
\begin{equation}
\int d^4x\; \left( J_{0}O_{0} +\eta_0 v_0^2 \right) \;. \label{lj}
\end{equation}
Therefore, for the one-loop contributions with a single external leg $J$ we get, upon proper expansion,
\begin{equation}
{\tilde J}(0) \left( \frac{1}{2} \chi(m_h^{2}) + \frac{(\delta\sigma)^{\left(1\right)}}{\lambda} v^2 -  \frac{ (\delta a)^{\left(1\right)}}{\lambda} v^2  \right) =0 \;, \label{zerO}
\end{equation}
which vanishes exactly, due  to equation (\ref{da1}). The quantity $\tilde J$ is the Fourier transformation of $J$. We see thus that fixing the vacuum counterterm
$(\delta a)^{\left(1\right)}$ as in eq.~(\ref{da1}), together with the requirement of vanishing tadpoles as in eq.~\eqref{tdp3}, yields automatically a vanishing condensate $\langle O\rangle_{pert}$.

Let us turn now to the Ward identity (\ref{hequation}), written at the quantum level in terms of the $1PI$ generator $\Gamma$, i.e.
\begin{eqnarray}
\int d^{4}x\left(\frac{\delta\Gamma}{\delta h}-\lambda v\frac{\delta\Gamma}{\delta J}\right)-\frac{\partial\Gamma}{\partial v} & = & \int d^{4}xv\left(J-2\eta\right)\;. \label{hequationgg}
\end{eqnarray}
Setting all fields and sources to zero and making use of $\langle O\rangle_{pert}=0$, it follows that equation (\ref{hequationgg}) gives nothing but
\begin{equation}
\frac{\partial {\cal E}_{v} }{\partial v} = \langle h\rangle -\lambda v \braket{O}\;. \label{wev}
\end{equation}
 To our knowledge, this is the first time in which the relation between the condition of the vanishing of the tadpoles, $\langle h \rangle = 0$,  and the minimization procedure of the vacuum energy ${\cal E}_v$ can be established in a direct way by means of a Ward identity, eq.~(\ref{hequationgg}), whose origin relies on the introduction from the beginning of the two local composite operators
$(O(x), V_\mu(x))$.  Notice that \eqref{wev} goes a bit beyond the standard textbook version, see e.g.~\cite{Peskin:1995ev} of splitting a quantum field as $\phi=\phi_{cl}+\eta$, where $\phi_{cl}$ is a ``classical'' background and $\eta$ the fluctuation, where it is understood that $\braket{\phi}= \phi_{cl}$ holds order per order with $\frac{\partial V}{ \partial \phi_{cl}}=0$ (minimization of the vacuum energy aka.~effective potential) if the tadpoles vanish, viz.~$\braket{\eta}=0$. In the current setting, we have established that BRST invariance allows for a priori two independent counterterms, one of which is related to the tadpoles and one to the vacuum energy minimization. It is in principle allowed to choose them differently, in return this will lead to a compensating $\braket{O}\neq 0$ so that \eqref{wev} is fulfilled.

 Noteworthy, eq.~\eqref{wev} is an exact functional constraint as directly arising from an underlying invariance of the theory, which validity in principle goes beyond perturbation theory. As such, even if $\braket{O}_{pert}=0$ order per order so that, perturbatively, the common equivalence between minimal vacuum energy and vanishing tadpoles is guaranteed, non-perturbative effects potentially leading to $\braket{O}_{non-pert}\neq0$ will then also alter the relation between the tadpoles and minimization condition, albeit in a constrained manner, namely according to \eqref{wev}. In the current Abelian case this might not happen, but it is more likely to occur when the generalization to the non-Abelian case will be considered in future work.

\section{The two-point current-current correlation function $\langle V_\mu(p) V_\nu(-p)\rangle $}\label{VV}

In this section we shall show that the transverse part of the the two-point correlation function of the composite operator $V_\mu$, i.e.~$\left\langle V_{\mu}\left(p\right)V_{\nu}\left(-p\right)\right\rangle ^{\textrm{transv}}$,
can be expressed exactly in terms of the two-point elementary Green function  $\left\langle A_{\mu}\left(p\right)A_{\nu}\left(-p\right)\right\rangle $, a result which is a direct consequence of the Ward identity (\ref{Widom}).  Moreover, the same Ward identity implies that  the longitudinal component, i.e.~$\left\langle V_{\mu}\left(p\right)V_{\nu}\left(-p\right)\right\rangle ^{\textrm{long}}$,
does not receive any quantum  correction beyond the tree level one, given by a momentum independent expression, up to a constant shift proportional to $\braket{O}$.  In particular, this last result, also exactly valid, implies that $\left\langle V_{\mu}\left(p\right)V_{\nu}\left(-p\right)\right\rangle ^{\textrm{long}}$ does not correspond to any propagating physical mode. This is an important consistency check of the gauge invariant description of the massive gauge vector particle in terms of the composite vector operator $V_\mu$. Only the transverse two-point function $\left\langle V_{\mu}\left(p\right)V_{\nu}\left(-p\right)\right\rangle ^{\textrm{transv}}$ describes a propagating excitation, corresponding to the 3 physical degrees of freedom of an observable massive vector particle.

Let us start thus by reminding the Ward identity  (\ref{Widom}), namely
\begin{eqnarray}
\frac{\delta\Gamma}{\delta A_{\mu}\left(x\right)}-2e\frac{\delta\Gamma}{\delta\Omega_{\mu}\left(x\right)}-e\Omega_{\mu}\left(x\right)\frac{\delta\Gamma}{\delta J\left(x\right)} & = & -\left(\partial^{2}\delta_{\mu\alpha}-\partial_{\mu}\partial_{\alpha}\right)A_{\alpha}\left(x\right)-i\partial_{\mu}b\left(x\right)+\frac{ev^{2}}{2}\Omega_{\mu}\left(x\right)\nonumber \\
&  & +\partial^{2}\Upsilon_{\mu}\left(x\right)-\partial_{\mu}\partial_{\nu}\Upsilon_{\nu}\left(x\right) \;,
\end{eqnarray}
which we rewrite in terms  of the generating  functional  $Z^c$ of the connected Green's
functions:
\begin{equation}
Z^c = \Gamma + \sum_{\rm fields \; \phi} \int d^4x \;J_{\phi} \phi \;, \label{wg}
 \end{equation}
yielding
\begin{eqnarray}
-J_{\mu}^{A}\left(x\right)-2e\frac{\delta Z^c}{\delta\Omega_{\mu}\left(x\right)}-e\Omega_{\mu}\frac{\delta Z^c}{\delta J\left(x\right)} & = & -\left(\partial^{2}\delta_{\mu\alpha}-\partial_{\mu}\partial_{\alpha}\right)\frac{\delta Z^c}{\delta J_{\alpha}^{A}\left(x\right)}-i\partial_{\mu}\frac{\delta Z^c}{\delta J^{b}\left(x\right)}+\frac{ev^{2}}{2}\Omega_{\mu}\left(x\right)\nonumber \\
&  & +\partial^{2}\Upsilon_{\mu}\left(x\right)-\partial_{\mu}\partial_{\nu}\Upsilon_{\nu}\left(x\right) \;,\label{eq:ward_v-1}
\end{eqnarray}
where $J_{\mu}^{A}$ and $J^{b}$ are the sources of $A_{\mu}$ and
$b$, respectively. Acting with  $\delta/\delta J_{\nu}^{A}\left(y\right)$
on  eq.~(\ref{eq:ward_v-1}) and taking all sources equal to zero, we get
\begin{eqnarray}
-\delta_{\mu\nu}\delta^4\left(x-y\right)+2e\left\langle V_{\mu}\left(x\right)A_{\nu}\left(y\right)\right\rangle  & = & \left(({\partial^2})^{x}\delta_{\mu\alpha}-\partial_{\mu}^{x}\partial_{\alpha}^{x}\right)\left\langle A_{\alpha}\left(x\right)A_{\nu}\left(y\right)\right\rangle +i\partial_{\mu}^{x}\left\langle b\left(x\right)A_{\nu}\left(y\right)\right\rangle \;, \label{ve1}
\end{eqnarray}
where the connected Green functions $ \langle V_{\mu}(x)V_{\nu}(y)\rangle $ and $\langle A_{\mu}(x)V_{\nu}(y)\rangle $ are given by
\begin{equation}
\langle V_{\mu}(x)V_{\nu}(y)\rangle = -\frac{\delta^2 Z^c}{\delta \Omega_\mu(x) \delta \Omega_\nu(y) } \Big|_{\rm sources=0} \;, \qquad \langle A_{\mu}(x)V_{\nu}(y)\rangle = -\frac{\delta^2 Z^c}{\delta J^A_\mu(x) \delta \Omega_\nu(y)} \Big|_{\rm sources=0} \;. \label{cgf}
\end{equation}
Employing now the Ward identity (\ref{eq:bequation}),  one can easily show the exact result
\begin{eqnarray}
\left\langle b\left(x\right)A_{\nu}\left(y\right)\right\rangle  & = & -\int\frac{d^{4}k}{\left(2\pi\right)^{4}}\frac{k_{\nu}}{k^{2}}e^{-ik\cdot\left(x-y\right)}
\end{eqnarray}
as well as the transversality of the elementary two-point correlation function  $\left\langle A_{\mu}\left(x\right)A_{\nu}\left(y\right)\right\rangle $, due to the Landau gauge $\partial A=0$. As a consequence, eq.~(\ref{ve1}) becomes
\begin{eqnarray}
2e\left\langle V_{\mu}\left(x\right)A_{\nu}\left(y\right)\right\rangle  & = & \partial^2_{x} \left\langle A_{\mu}\left(x\right)A_{\nu}\left(y\right)\right\rangle +\left(\delta_{\mu\nu}-\frac{\partial_{\mu}\partial_{\nu}}{\partial^{2}}\right)^{xz}\delta^4\left(z-y\right),\label{eq:va_aa-1}
\end{eqnarray}
or, equivalently, in momentum space
\begin{eqnarray}
2e\left\langle V_{\mu}\left(p\right)A_{\nu}\left(-p\right)\right\rangle  & = & -p^{2}\left\langle A_{\mu}\left(p\right)A_{\nu}\left(-p\right)\right\rangle +\mathcal{P}_{\mu\nu}\left(p\right) \;, \label{ve2}
\end{eqnarray}
To proceed, we act with  $\delta/\delta\Omega_{\nu}\left(y\right)$ on eq.~(\ref{eq:ward_v-1}), obtaining a relationship between $\left\langle V_{\mu}\left(x\right)A_{\nu}\left(y\right)\right\rangle $
and $\left\langle V_{\mu}\left(x\right)V_{\nu}\left(y\right)\right\rangle $, i.e.
\begin{eqnarray}
	 2e\left\langle V_{\mu}\left(x\right)V_{\nu}\left(y\right)\right\rangle -e\delta_{\mu\nu}\delta^4\left(x-y\right)\left\langle O\left(x\right)\right\rangle  & = & \left(\partial^{2}\delta_{\mu\alpha}-\partial_{\mu}\partial_{\alpha}\right)^{x}\left\langle A_{\alpha}\left(x\right)V_{\nu}\left(y\right)\right\rangle +i\partial_{\mu}^{x}\left\langle b\left(x\right)V_{\nu}\left(y\right)\right\rangle \nonumber\\&&+\frac{ev^{2}}{2}\delta_{\mu\nu}\delta^4\left(x-y\right)	\;. \label{ve3}
\end{eqnarray}
The equation above can be greatly simplified since from the BRST invariance of $V_\mu$ it follows that
\begin{equation}
\left\langle b\left(x\right)V_{\nu}\left(y\right)\right\rangle = \langle s({-i\bar c}(x) V_\nu(y) ) \rangle = 0 \;.
\end{equation}
Employing then eq.~(\ref{eq:va_aa-1}), we find
\begin{eqnarray}
4e^{2}\left\langle V_{\mu}\left(x\right)V_{\nu}\left(y\right)\right\rangle  & = & ({\partial}^4)^{x}\left\langle A_{\mu}\left(x\right)A_{\nu}\left(y\right)\right\rangle +({\partial}^2)^{x}\left(\delta_{\mu\nu}-\frac{\partial_{\mu}\partial_{\nu}}{\partial^{2}}\right)^{xz}\delta^4\left(z-y\right)+(m^{2}+2e^2\braket{O})\delta_{\mu\nu}\delta^4\left(x-y\right)\nonumber \\ \;, \label{ve4}
\end{eqnarray}
or, in momentum space:
\begin{eqnarray}
4e^{2}\left\langle V_{\mu}\left(p\right)V_{\nu}\left(-p\right)\right\rangle  & = & p^{4}\left\langle A_{\mu}\left(p\right)A_{\nu}\left(-p\right)\right\rangle -p^{2}\mathcal{P}_{\mu\nu}\left(p\right)+(m^{2}+2e^2\braket{O})\delta_{\mu\nu} \;. \label{ve5}
\end{eqnarray}
Splitting now equation (\ref{ve5}) into transverse and longitudinal components, we get the announced exact results:
\begin{eqnarray}
\mathcal{P}_{\mu\nu}(p) \left\langle V_{\mu}\left(p\right)V_{\nu}\left(-p\right)\right\rangle & = & \frac{p^{4}}{4e^{2}}\mathcal{P}_{\mu\nu}\left(p\right)\left\langle A_{\mu}\left(p\right)A_{\nu}\left(-p\right)\right\rangle -3\frac{\left(p^{2}-m^{2}-2e^2\braket{O}\right)}{4e^{2}}\label{eq:transversal_result} \;,
\end{eqnarray}
and
\begin{eqnarray}
\mathcal{L}_{\mu\nu}(p)\left\langle V_{\mu}\left(p\right)V_{\nu}\left(-p\right)\right\rangle & = & \frac{v^{2}}{4}+ \frac{\braket{O}}{2}\label{eq:longitudinal_result} \;,
\end{eqnarray}
Eqs.~(\ref{eq:transversal_result}) and (\ref{eq:longitudinal_result}) show in fact that $\left\langle V_{\mu}\left(p\right)V_{\nu}\left(-p\right)\right\rangle ^{\textrm{transv}}$ can be  expressed in terms of $\left\langle A_{\mu}\left(p\right)A_{\nu}\left(-p\right)\right\rangle$ and that $\left\langle V_{\mu}\left(p\right)V_{\nu}\left(-p\right)\right\rangle ^{\textrm{long}}$ does not receive any quantum correction up to perhaps a constant shift in $\braket{O}$.

In Appendix~\ref{A} one finds a detailed one-loop verification of the  results (\ref{eq:transversal_result}) and (\ref{eq:longitudinal_result}). To that end, let us end this section by rewriting eqs.~(\ref{eq:transversal_result}),(\ref{eq:longitudinal_result}) in a slightly different form which will turn out to be helpful for the one-loop check. Adopting as before a renormalization consistent with vanishing tadpoles and $\frac{\partial \mathcal{E}_v}{\partial v}=0$, we already have proven that this implies that $\braket{O}_{pert}=0$. Looking at the expression for the vector operator $V_\mu(x)$, i.e.
\begin{equation}
V_{\mu}=\frac{v}{2}\left(evA_{\mu}+\partial_{\mu}\rho\right)\; +\ldots, \label{vmu}
\end{equation}
it is convenient to subtract from the correlator $\left\langle V_{\mu}\left(x\right)V_{\nu}\left(y\right)\right\rangle $ the quantities $\left\langle A_{\mu}\left(x\right)A_{\nu}\left(y\right)\right\rangle $
and $\left\langle \rho\left(x\right)\rho\left(y\right)\right\rangle $. Accordingly, we introduce the subtracted correlation function
\begin{eqnarray}
	\widetilde{\left\langle V_{\mu}\left(x\right)V_{\nu}\left(y\right)\right\rangle } & := & \left\langle V_{\mu}\left(x\right)V_{\nu}\left(y\right)\right\rangle -\frac{e^{2}v^{4}}{4}\left\langle A_{\mu}\left(x\right)A_{\nu}\left(y\right)\right\rangle -\frac{v^{2}}{4}\partial_{\mu}^{x}\partial_{\nu}^{y}\left\langle \rho\left(x\right)\rho\left(y\right)\right\rangle \;. \label{subt}
\end{eqnarray}
From eqs.~(\ref{eq:transversal_result}) and (\ref{eq:longitudinal_result}) we get
\begin{eqnarray}
\mathcal{P}_{\mu\nu}(p) \widetilde{\left\langle V_{\mu}\left(p\right)V_{\nu}\left(-p\right)\right\rangle} & = & \frac{\left(p^{4}-m^{4}\right)}{4e^{2}}\mathcal{P}_{\mu\nu}\left(p\right)\left\langle A_{\mu}\left(p\right)A_{\nu}\left(-p\right)\right\rangle -3\frac{\left(p^{2}-m^{2}\right)}{4e^{2}} \;,\label{eq:transversal_result-1}
\end{eqnarray}
and
\begin{eqnarray}
\mathcal{L}_{\mu\nu}(p) \widetilde{\left\langle V_{\mu}\left(p\right)V_{\nu}\left(-p\right)\right\rangle } & = & \frac{v^{2}}{4}-\frac{v^{2}p^{2}}{4}\left\langle \rho\left(p\right)\rho\left(-p\right)\right\rangle \label{eq:longitudinal_result-1} \;,
\end{eqnarray}
meaning that
\begin{eqnarray}
\widetilde{\left\langle V_{\mu}\left(p\right)V_{\nu}\left(-p\right)\right\rangle }_{\textrm{$n$-loop}}^{\textrm{transv}} & = & \frac{\left(p^{4}-m^{4}\right)}{4e^{2}}\mathcal{P}_{\mu\nu}\left(p\right)\left\langle A_{\mu}\left(p\right)A_{\nu}\left(-p\right)\right\rangle _{\textrm{$n$-loop}} \;, \qquad n\ge 1 \;,\label{eq:transversal_result-1-1}\\
\widetilde{\left\langle V_{\mu}\left(p\right)V_{\nu}\left(-p\right)\right\rangle }_{\textrm{$n$-loop}}^{\textrm{long}} & = & -\frac{v^{2}p^{2}}{4}\left\langle \rho\left(p\right)\rho\left(-p\right)\right\rangle _{\textrm{$n$-loop}} \;, \qquad n\ge 1 \;.\label{eq:longitudinal_result-1-1}
\end{eqnarray}

\section{The Higgs model revisited by means of the Equivalence Theorem}\label{EQ}
\subsection{Step 1: from cartesian to polar coordinates}
In order to investigate the field redefinition allowing to move from cartesian to polar coordinates within the Equivalence Theorem \cite{Bergere:1975tr,Haag:1958vt,Kamefuchi:1961sb,Lam:1973qa}, let us consider the starting partition function of the Higgs model, i.e.
\begin{equation}
{\cal Z}_{\rm Higgs} = \int [D\Phi] \; e^{-\left( S_{\rm Higgs}(A,h, \rho) + \int d^4x \left (i b\partial A +{\bar c}\partial^2 c\right) \right)} \;, \label{Zh}
\end{equation}
with $S_{\rm Higgs}(A,h, \rho)$ given in expressions (\ref{ah2}) equipped with the cartesian parametrization of eq.~(\ref{exp}). The measure $[D\Phi]$ denotes integration over all fields $(A_\mu, h, \rho, b, {\bar c}, c)$.

Let us perform the field transformation enabling us to move to the polar parametrization $(h', \rho', A'_\mu)$, namely
\begin{eqnarray}
h & \rightarrow & (h' + v) \cos(\rho') - v \;, \nonumber \\
\rho & \rightarrow & (h' + v) \sin(\rho') \;, \nonumber \\
A_\mu & \rightarrow &  A'_\mu - \frac{1}{e} \partial_\mu \rho' \;, \label{pc}
\end{eqnarray}
the remaining fields $(b,{\bar c},c)$ being left unchanged. For the partition function ${\cal Z}_{\rm Higgs} $ we get thus
\begin{equation}
{\cal Z}_{\rm Higgs} = \int [D\Phi'] \; \left( \det {\cal M} \right) \; e^{-\left( S_{\rm Higgs}(A',h', \rho') + \int d^4x \left (i b(\partial A' -\frac{1}{e} \partial^2 \rho' ) +{\bar c}\partial^2 c\right) \right)} \;, \label{Zhp}
\end{equation}
where $[D\Phi'] = [DA' \;Dh' \;D\rho' \;Db \;D{\bar c} \;Dc]$ and $(\det {\cal M})$ stands for the Jacobian stemming from (\ref{pc}), i.e.
\begin{equation}
(\det {\cal M}) = \det \left( \delta^4(x-y)
\left( \begin{array}{ccc}
\cos(\rho'(x)) & -(h'(x)+v) \sin(\rho'(x))  & 0 \\
 \sin(\rho'(x)) & (h'(x)+v) \cos(\rho'(x)) & 0 \\
0 & -\frac{1}{e}\partial_\nu^x & \delta_{\nu\mu}
\end{array} \right)  \right)\;.  \label{det}
\end{equation}
As already mentioned, the $(\det {\cal M}) $ can be exponentiated by introducing a suitable set of anti-ghosts $({\bar \eta}, {\bar \sigma}, {\bar \xi}_\nu)$ as well as a set of ghosts $(\eta, \sigma,\xi_\nu)$, so that
\begin{equation}
(\det {\cal M}) = \int [D(\rm {new \; ghosts})] \; e^{-S_{\rm ghosts, 1}} \;, \label{sgh}
\end{equation}
with $ [D{\rm ({new \; ghosts})}] = [D {\bar \eta} \;D {\bar \sigma} \; D {\bar \xi} \; D\eta \; D \sigma \; D\xi ]$ and
\begin{eqnarray}
S_{\rm ghosts, 1}  & = & \int d^4x \left( {\bar \eta} \eta \cos(\rho') - {\bar \eta} \sigma (h' + v) \sin(\rho') - \frac{1}{e} {\bar \xi}_\nu \partial_\nu \sigma \right) \nonumber \\
& \; \; + & \int d^4x \left( {\bar \sigma} \eta \sin(\rho') + {\bar \sigma} \sigma (h'+ v) \cos(\rho')  + {\bar \xi}_\nu \xi_\nu \right) \;. \label{ngh}
\end{eqnarray}
Therefore, for the partition function we get
\begin{equation}
{\cal Z}_{\rm Higgs} = \int [D\Phi'] [D(\rm {new \; ghosts})]\;  e^{- S_{\rm eff} } \;, \label{Zhp1}
\end{equation}
with
\begin{eqnarray}
S_{\rm eff} &=& S_0+ S_{\rm ghosts, 1} \;, \nonumber\label{eff}\\
S_0(A',h', \rho')&=&S_{\rm Higgs}(A',h', \rho') + \int d^4x \left (i b (\partial A' -\frac{1}{e} \partial^2 \rho' ) +{\bar c}\partial^2 c\right)
\end{eqnarray}
It is easy to check now that expression (\ref{eff}) is left invariant by the following nilpotent BRST transformations
\begin{eqnarray}
s  A'_\mu & = & 0 \;, \nonumber \\
s h' & = & 0 \;, \nonumber \\
s \rho' & = & e c \;, \qquad  sc = 0 \;, \nonumber \\
s {\bar c} & = & ib \;, \qquad s b=0 \;, \nonumber \\
s {\bar \xi}_\nu & = & s \xi_\nu = s \eta = s \sigma = 0 \;, \nonumber \\
s {\bar \eta} & = & - e c {\bar \sigma} \;, \nonumber \\
s {\bar \sigma} & = & e c {\bar \eta} \;, \label{nbrst}
\end{eqnarray}
with
\begin{equation}
s S_{\rm eff} = 0 \;, \qquad s^2 =0 \;. \label{ninv}
\end{equation}
As pointed out in \cite{Blasi:1998ph}, the existence of the BRST transformations  (\ref{nbrst}) guarantees that the field redefinition (\ref{pc}) is harmless for physical quantities, as stated by the Equivalence Theorem \cite{Bergere:1975tr,Haag:1958vt,Kamefuchi:1961sb,Lam:1973qa}. The new ghosts $({\bar \eta}, {\bar \sigma}, {\bar \xi}_\nu)$ and  $(\eta, \sigma,\xi_\nu)$ turn out to compensate \cite{Blasi:1998ph} the effects of the change of variable (\ref{pc}).

Moreover, taking a look at the Higgs action $S_{\rm Higgs}(A',h',\rho') $ in terms of the new field variables, one gets
\begin{equation}
S_{\rm Higgs}(A',h',\rho') = \int d^4x \left(  \frac{1}{4} F^2(A') + \frac{1}{2} (\partial_\mu h' )^2          + \frac{e^2}{2} (A')^2 (v+h')^2 +\frac{\lambda}{8} \left( {h'}^2 + 2 h' v \right)^2 \right) \;, \label{nsh}
\end{equation}
where, according to the BRST transformations (\ref{nbrst}), the new variables $(A'_\mu, h')$ are gauge invariant. Said otherwise, there is no more trace in expression (\ref{nsh}) of the original $U(1)$ local gauge invariance, as everything is expressed in terms of the gauge invariant fields $(A'_\mu, h')$. From eq.~(\ref{pc}) one notices in fact that the Goldstone boson $\rho'$ has been eaten by the vector field, expressing the physical content of the Higgs phenomenon. This renowed  feature of the polar parametrization is nicely captured by the BRST transformations (\ref{nbrst}), from which one observes that the fields $(\rho',c)$ form a so-called BRST doublet \cite{Piguet:1995er}, i.e.
\begin{equation}
s \rho'  =  e c \;, \qquad  sc = 0 \;. \label{dbb}
\end{equation}
From the general results on the BRST cohomology \cite{Piguet:1995er}, it follows that these fields do not contribute to the non-trivial cohomology of the BRST operator. As such, the field $\rho'$ cannot enter the explicit expressions of the BRST  invariant physical operators, see eqs.~(\ref{opn}) below.

\subsection{Step 2: from polar coordinates to the gauge invariant operators $(O, V_\mu)$}
 We can further improve upon the previous Step 1, not only to formally banish the new ghosts $({\bar \eta}, {\bar \sigma}, {\bar \xi}_\nu),\eta, \sigma,\xi_\nu)$ to the unphysical sector, but to actually show the extra ghost piece of the action, $S_{\rm ghost}$, is akin to a gauge fixing.

We first notice that for the  gauge invariant composite operators $(O, V_\mu)$ in the new variables $(A'_\mu, h')$, one obtains
\begin{equation}
O(A'_\mu, h') =\frac{1}{2} (h' + v)^2-\frac{v^2}{2} \;, \qquad V_\mu(A', h') = \frac{e}{2} (h' + v)^2 A'_\mu \;. \label{opn}
\end{equation}
As expected, no dependence from the Goldstone boson $\rho' $ is found here. The foregoing relations \eqref{opn} can be inverted as follows
\begin{equation}
h' = \frac{O}{v^2}\left(1+\zeta f_1(O/{v^2})\right)\;,\qquad A'_\mu = \frac{2 V_\mu}{ev^2}\left(1+\zeta f_2(O/v^2)\right) \label{opn2}
\end{equation}
where we introduced a new parameter $\zeta$ in front of the non-linear part, encoded in the quantities $f_1$ and $f_2$, which are power series in $(O/{v^2})$. Its role will become clear soon.

Next, we consider \eqref{opn2} as a path integral field transformation. As before, we may introduce a set of ghosts $(\omega, \bar\omega, \omega_\mu,\bar\omega_\mu)$  to exponentiate the corresponding Jacobian. One arrives at a new\footnote{And at first sight non-renormalizable because of the vertices containing inverse powers of $v$.} classical action
\begin{equation}\label{opn3}
  S_{\rm new}(V_\mu, O, \rho')= S_0(2V_\mu/(ev^2),O/v^2,\rho')+ \zeta S_{1}(V_\mu, O, \rho')+S_{\rm ghosts, 1} + S_{\rm ghosts, 2}
  \end{equation}
where
\begin{equation}\label{opn4}
  \zeta S_{1}(V_\mu, O, \rho')= S_{0}(2 V_\mu/(ev^2)\left(1+\zeta f_2(O/v^2)\right), \frac{O}{v^2}\left(1+\zeta f_1(O/{v^2})\right),\rho') - S_{0}(2V_\mu/(ev^2),O/v^2,\rho')
\end{equation}
and
\begin{equation}\label{opn5}
  S_{\rm ghosts, 2}=\int d^4x\left(
                 \begin{array}{cc}
                   \bar\omega & \bar\omega_\mu \\
                 \end{array}
               \right)\left(
                        \begin{array}{cc}
                          \frac{1}{v^2}+\zeta \frac{\delta f_1}{\delta O} & 0 \\
                          \zeta V_\mu \frac{\delta f_2}{\delta O} & \frac{2}{ev^2}\delta_{\mu\nu}(1+\zeta f_1) \\
                        \end{array}
                      \right)\left(
                               \begin{array}{c}
                                  \omega \\
                                 \omega_\nu \\
                               \end{array}
                             \right).
\end{equation}
Given the gauge invariant nature of both $O$ and $V_\mu$, the BRST transformation $s$ can be naturally generalized to the new ghosts,
\begin{equation}\label{opn6}
  s\omega = s\bar\omega= s\omega_\mu = s\bar\omega_\mu=0.
\end{equation}
On top of the BRST $s$, we now introduce another nilpotent (Grassmann) symmetry generator $\delta$,
\begin{eqnarray}
  \delta B_\eta  &=& \bar\eta\,,\qquad \delta B_\sigma = \bar\sigma\,,\qquad \delta B_{\xi,\mu}= \bar\xi_\mu\,, \nonumber\\
  \delta B_\omega &=& \bar\omega\,, \qquad\delta B_{\omega,\mu} = \bar\omega_\mu\,,\nonumber \\
  \delta\beta &=& \zeta\,,\qquad \delta \zeta=0\,,\nonumber\\
  \delta(\rm rest)&=&0,
\end{eqnarray}
where, following \cite{Blasi:1998ph}, we introduced new local ghosts $(B_\eta, B_\sigma, B_{\xi,\mu}, B_\omega, B_{\omega, \mu})$ and a global ghost $\beta$, left invariant by the BRST operator, namely
\begin{equation}
s\; (B_\eta, B_\sigma, B_{\xi,\mu}, B_\omega, B_{\omega, \mu}, \beta, \zeta) = 0 \;, \label{bnwf}
\end{equation}
from which it follows that
\begin{equation}
\delta^2=0 \;, \qquad \{s,\delta\}=0 \;. \label{twon}
\end{equation}
For convenience, we can replace
\begin{eqnarray}
S_{\rm ghosts, 1}  & \to & S_{\rm ghosts, 1}= \zeta\int d^4x \left( {\bar \eta} \eta \cos(\rho') - {\bar \eta} \sigma (h' + v) \sin(\rho') - \frac{1}{e} {\bar \xi}_\nu \partial_\nu \sigma \right) \nonumber \\
& \; \; + & \zeta\int d^4x \left( {\bar \sigma} \eta \sin(\rho') + {\bar \sigma} \sigma (h'+ v) \cos(\rho')  + {\bar \xi}_\nu \xi_\nu \right) \;. \label{ngh}
\end{eqnarray}
which corresponds to pulling out a factor of $\sqrt\zeta$ from each of the fields $({\bar \eta}, {\bar \sigma}, {\bar \xi}_\nu,\eta, \sigma,\xi_\nu)$. Importantly, this does not alter the BRST variations \eqref{nbrst}. We can then rewrite the new action \eqref{opn3} as
\begin{equation}\label{opn10}
  S_{\rm new}= S_{0}(2V_\mu/(ev^2),O/v^2,\rho')+ \delta(\beta S_{1})+\delta \mathcal{G}_1 + \delta\mathcal{G}_2
  \end{equation}
where
\begin{eqnarray}\label{opn11}
  \mathcal{G}_1 &=&\zeta\int d^4x \left( {B_\eta} \eta \cos(\rho') - {B_\eta} \sigma (h' + v) \sin(\rho') - \frac{1}{e} B_{\xi,\nu} \partial_\nu \sigma \right) \nonumber \\
& \; \; + & \zeta\int d^4x \left( {B_\sigma} \eta \sin(\rho') + {B_\sigma} \sigma (h'+ v) \cos(\rho')  + B_{\xi,\nu} \xi_\nu \right)\,,\nonumber\\ \mathcal{G}_2&=&\int d^4x\left(
                 \begin{array}{cc}
                   B_\omega & B_{\omega,\mu} \\
                 \end{array}
               \right)\left(
                        \begin{array}{cc}
                          \frac{1}{v^2}+\zeta \frac{\delta f_1}{\delta O} & 0 \\
                          \zeta V_\mu \frac{\delta f_2}{\delta O} & \frac{2}{ev^2}\delta_{\mu\nu}(1+\zeta f_1) \\
                        \end{array}
                      \right)\left(
                               \begin{array}{c}
                                  \omega \\
                                 \omega_\nu \\
                               \end{array}
                             \right).
  \end{eqnarray}
 Equivalently, we could have introduced an extended BRST invariance \cite{Delduc:1996yh} corresponding to the generalized  nilpotent operator $\tilde s=s+\delta$, with $\tilde s S_{\rm new}=0$, and derive everything from there, although in the current paper, we will
immediately work at the level of\footnote{Both formulations are equivalent upon introducing a proper filtration, see \cite{Delduc:1996yh}.} $s$ and $\delta$.

We will also need a new ghost charge, defined as $+1$ for $(\sigma,\eta,\xi_\mu,\omega,\omega_\mu)$, $-1$ for $(\bar\sigma,\bar\eta,\bar\xi_\mu,\bar\omega,\bar\omega_\mu)$ and $-2$ for $(B_\eta, B_\sigma, B_{\xi,\mu}, B_\omega, B_{\omega, \mu},\beta)$. The other fields remain uncharged.
The operator $\delta$ then increases this charge by one unit. Clearly, the action $S_{\rm new}$ carries no such charge.

To define the physical subspace, we can now first identify the BRST $s$-cohomology, which contains the (standard) gauge invariant operators of the Abelian Higgs model, supplemented with $s$-invariant operators constructed from the new ghosts which are not $s$-exact. To remove these extra operators from the physical subspace, we can further restrict within that
$s$-cohomology to those field functionals that belong to the $\delta$-cohomology. This is an example of a constrained cohomology, a concept which was already successfully employed in other cases as, for example,   the characterization  of  the observables of topological Yang-Mills theory \cite{Delduc:1996yh}, see also \cite{Ouvry:1988mm,Stora:1996yc}. In our case, since \emph{all} the newly introduced ghost fields during Step 1 as well as Step 2 form $\delta$-doublets, this constrained cohomology will just contain the original gauge invariant operators, as desired. Moreover, these operators can be re-expressed in terms of
 $O$ and $V_\mu$.

The dependence of the quantum effective action $\Gamma$, and thus of the $\cal S$-matrix, on the parameter $\zeta$ is well under-control, and can also be expressed in a neat functional way.
The functional  $\delta$-operator reads
\begin{equation}\label{funct1}
  \mathcal{D} =\int d^4x \left( \bar\eta\frac{\delta}{\delta B_\eta} + \bar\sigma\frac{\delta}{\delta B_\sigma}+\bar\xi_{\mu}\frac{\delta}{\delta B_{\xi,\mu}}+\bar\omega\frac{\delta}{\delta B_\omega}+\bar\omega_\mu\frac{\delta}{\delta B_{\omega,\mu}}+\zeta \frac{\partial}{\partial \beta}\right),\qquad \mathcal{D}^2=0\,,
\end{equation}
and since it is linear, it can be used at the quantum level.

Evidently, we have $\mathcal{D} S_{\rm new}=0$, which for the quantum effective action $\Gamma$ implies
\begin{equation}\label{funct2}
  \mathcal{D}\Gamma=0,
\end{equation}
which is non-anomalous since there is no room for $\delta$-non-cohomologically trivial breaking terms, as the new ghost charged fields are all coming in $\delta$-doublets.

Moreover, for any functional $\mathcal{F}$, it can be verified that
\begin{equation}\label{funct3}
  \frac{\partial}{\partial\zeta} \mathcal{D}\mathcal{F}-\mathcal{D}\frac{\partial\mathcal{F}}{\partial\zeta}=\frac{\partial \mathcal F}{\partial \beta},
\end{equation}
which we can use to show that
\begin{eqnarray}  \label{funct4}
  \frac{\partial S_{\rm new}}{\partial \zeta} &=& \frac{\partial}{\partial \zeta}\mathcal{D}\mathcal{Y}  = \mathcal{D}\frac{\partial\mathcal Y}{\partial\zeta}+\frac{\partial \mathcal Y}{\partial \beta}
   = \mathcal D\frac{\partial \mathcal Y}{\partial \zeta}  +S_1.
\end{eqnarray}
We introduced the shorthand
\begin{equation}\label{shorthand}
  \mathcal{Y}=\beta S_{1}+ \mathcal{G}_1 + \mathcal{G}_2.
\end{equation}
Unfortunately, the r.h.s.~of \eqref{funct4} is not $\mathcal{D}$-exact, but fortunately
\begin{eqnarray}  \label{funct5}
  \zeta\frac{\partial S_{\rm new}}{\partial \zeta} &=& \mathcal D\left(\zeta\frac{\partial \mathcal Y}{\partial \zeta}  +\beta S_1\right)
\end{eqnarray}
is. Thanks to the Quantum Action Principle, \cite{Piguet:1995er}, this classical result can be extended to the quantum level, i.e.
\begin{eqnarray}  \label{funct6}
  \zeta\frac{\partial \Gamma}{\partial \zeta} &=& \mathcal D\left( \Delta_{\mathcal{Y}}\cdot\Gamma  + \Delta_1\cdot\Gamma\right)
\end{eqnarray}
where $\Delta_{\mathcal{Y}}\cdot\Gamma=\zeta\frac{\partial \mathcal Y}{\partial \zeta}+\mathcal{O}(\hbar)$ and  $\Delta_1\cdot\Gamma=\beta S_1+\mathcal{O}(\hbar)$  are quantum insertions reducing to the operators present in \eqref{funct5} for $\hbar\to0$ \cite{Piguet:1995er}. Their precise form is of no interest to us, as the final
relation \eqref{funct6} is sufficient to conclude that the $\zeta$-dependent pieces of the quantum action are relegated to a cohomologically trivial sector and as such will have no bearing on physical expectation values, that is, the observables.
A fortiori, the dangerous ``non-renormalizable'' terms stemming from $f_1$ and $f_2$ in \eqref{opn2} will not lead to non-curable UV divergences in physical correlation functions.

The foregoing analysis implies that at the end, the physically relevant piece of the action can be expressed in terms of the gauge invariant operators $O$ and $V_\mu$,
where only the leading (linear) terms in \eqref{opn2} are relevant for the physics. Concretely, this amounts to considering
\begin{eqnarray}\label{finalaction}
  S_{\rm new}&=& \int d^4x \left(  \frac{1}{4} F^2(2V_\mu/(ev^2)) + \frac{1}{2v^4} (\partial_\mu O )^2   + \frac{2}{v^2}V_\mu^2       + \frac{4}{v^4} V_\mu^2 O +\frac{\lambda}{2}O^2 \right)
  \nonumber\\
  &+&\int d^4x \left( \frac{2ib}{ev^2}\partial_\mu V_\mu\underbrace{- \frac{ib}{e}\partial^2 \rho'+\bar c \partial^2 c}_{=-\frac{1}{e}s(\bar c \partial^2 \rho')}\right)+\delta{\rm-cohomologically~irrelevant~pieces}.
\end{eqnarray}
Notice that the underbraced part contains a physically irrelevant piece, being $s$-exact, which is actually a remnant of the original (Landau) gauge fixing, while the other piece assures that $V_\mu$ will be transverse on-shell (as expected).
The physical correlation functions are fully determined by the (renormalizable) vertices and propagators derivable by \eqref{finalaction}.

\subsection{On the $U(1)$ symmetry}
Once the action is rewritten in terms of the gauge invariant variables, one might wonder about the role of the (global) $U(1)$ and its symmetry breaking, since its role is diminished in the new formulation. Needless to say, $V_\mu$ is still the corresponding Noether current.   However, we no longer have $ \braket{\varphi}$ at our disposal to decide about the vacuum being (globally) invariant or not. At the end, this is not what interests us, but rather whether the vector particles are massive (or not). If a transition from massive to massless behaviour corresponds to an actual phase transition can be analyzed in terms of the analytic properties of the free energy.

Concretely, in the current BRST invariant framework, we can always introduce a gauge invariant parameter $v\neq0$ as the minimum of the classical potential and keep it as minimum for the quantum potential by a suitable choice of the vacuum renormalization constant\footnote{And depending on the value of $\braket{O}$, $\braket{\varphi}$ and $\frac{v}{\sqrt{2}}$ may coincide or not, depending on whether $\braket{h}$ is zero or not.}.
 Evidently, physics will not depend on this choice of renormalization scheme. Depending on its own dynamics, the theory will then tell us whether $\braket{O}$ remains nonzero or not, and whether the gauge invariant vector quantity $V_\mu$ keeps its nonzero mass pole beyond tree level.  This is in perfect accordance with the analysis presented in \cite{Frohlich:1980gj,Frohlich:1981yi}. In fact, it was shown in these references that in some classes of gauge,  $\braket{\varphi}=0$ due to (non-trivial) quantum effects. Evidently, this does not imply that the observable vector particles would not be massive in these gauges.

The rewriting of the action in terms of the explicitly gauge invariant variables, in conjunction with the Equivalence Theorem we worked out, is an explicit framework capable of implementing consistently the main message of \cite{Frohlich:1980gj,Frohlich:1981yi}, namely: we may choose around which value of $\bar\varphi$ on the $\varphi$-orbit ($v$ in our language) one expands, whilst standard perturbation theory as developed in textbooks corresponds to the  situation of picking up a particular direction, \textit{i.e.}~setting $\braket{\varphi}=\bar\varphi(=\frac{v}{\sqrt 2})$.  At the end, the observable physics will be the same, irrespective of any gauge choice or assumption about broken global $U(1)$ invariance.

\section{Conclusion}
In this work we have exploited the content of two new Ward identities, eqs.~(\ref{hequation}),(\ref{Widom}), which can be obtained when the two gauge invariant operators $(O(x),V_\mu(x))$, eq.~(\ref{ovop}), are introduced in the $U(1)$ Higgs model from the beginning. As already discussed in \cite{Dudal:2019aew,Dudal:2019pyg,Capri:2020ppe}, the renormalizable operators $(O(x),V_\mu(x))$ offer a truly gauge invariant framework for the description of the Higgs and gauge vector boson particles \cite{hooft2012nonperturbative,Frohlich:1980gj,Frohlich:1981yi}.

These additional Ward identities have far reaching consequences. For instance, the Ward identity (\ref{hequation}), corresponding to the inclusion of the scalar operator $O(x)$, has enabled us to connect in an explicit way the stationary condition for the vacuum energy ${\cal E}_v$ the vanishing of the tadpole diagrams and the vacuum condensate $\braket{O}$, as expressed by eq.~(\ref{Wennw}), see the discussion given in Section~\ref{OC}. To our knowledge, this is the first time in which such a relationship has been established by means of a Ward identity.

Concerning the vector operator $V_\mu(x)$, it turns out that it can be identified with the conserved Noether current of the global $U(1)$ invariance, eqs.~(\ref{gbo}), (\ref{cnst}), displayed by the action of the Higgs model. The second additional Ward identity (\ref{Widom}) can be seen in fact as the translation at the functional level of the conservation law obeyed by $V_\mu(x)$. Also here, the identity (\ref{Widom}) has deep consequences, see eqs.~(\ref{ta1}), (\ref{la2}). In particular, the transverse component of the two-point function $\langle V_\mu(p) V_\nu(-p)\rangle $ can be obtained exactly from the elementary correlator $\langle A_\mu(p) A_\nu(-p)\rangle $, a fact which ensures that the pole masses of both correlation functions are identical. Moreover, the longitudinal component of $\langle V_\mu(p) V_\nu(-p)\rangle $ does not receive any momentum dependent contribution, eq.~(\ref{la2}). As a consequence, the longitudinal part of $\langle V_\mu(p) V_\nu(-p)\rangle $ cannot describe a propagating mode. In the final Section~\ref{EQ}, we have in a first step discussed the connection between the cartesian and the polar parametrization of the complex scalar field $\varphi$ in the light  of the Equivalence Theorem \cite{Bergere:1975tr,Haag:1958vt,Kamefuchi:1961sb,Lam:1973qa}, reformulated within a BRST framework \cite{Blasi:1998ph}. The Jacobian relating the two parametrizations can be exponentiated by introducing a new set of ghost fields, leading to the BRST transformations displayed in
eq.~(\ref{nbrst}). We then introduced another set of fields and associated BRST transformation to show that the non-linear pieces in the field redefinitions are akin to a gauge fixing term and are as such irrelevant for physical observables. This generalization of the results \cite{Blasi:1998ph} leads to a constrained BRST cohomology characterization of the gauge invariant observables. These transformations guarantee that field redefinitions have no effects on physical quantities, according to the Equivalence Theorem \cite{Bergere:1975tr,Haag:1958vt,Kamefuchi:1961sb,Lam:1973qa}. This had lead to an explicitly gauge invariant action \eqref{finalaction} leading to renormalizable physical correlation functions and thus $\cal S$-matrix elements.

 Let us end by mentioning that we are now investigating to what extent all of these results can be generalized to the non-Abelian $SU(2)$ Higgs model with a single scalar field in the fundamental representation \cite{prep}, with as final aim developing a sensible fully gauge invariant study of the more complex electroweak theory $SU(2)_L \times U(1)$, as realized in Nature. Apart from that, even in the Abelian case some topics deserve further investigations, for example the inclusion of vortices, \cite{Nielsen:1973cs}, in the gauge invariant reformulation or the case where the parameter $v$ has a purely dynamical origin, \cite{Coleman:1973jx}. Notice that we can still define $v$ as the minimum of the potential, the connection with tadpoles and potential vacuum expectation value of $O$, see also \cite{Knecht:2001cc}, will still be encoded  in the identity \eqref{wev}.

\section*{Acknowledgements}
The authors would like to thank the Brazilian agencies CNPq and FAPERJ for financial support. This study was financed in part
by the Coordena{\c c}{\~a}o de Aperfei{\c c}oamento de Pessoal de N{\'\i}vel Superior--Brasil (CAPES) --Finance Code 001. S.P.~Sorella is a level $1$ CNPq researcher under the contract 301030/2019-7.

\vspace{3cm}

\appendix

\section{Appendix A: some explicit one-loop verifications \label{A} }

\subsection{Evaluation of the correlation function  $\widetilde{\left\langle V_{\mu}\left(x\right)V_{\nu}\left(y\right)\right\rangle }$}
At one-loop order, the diagrams contributing to  the two-point function of the vector operators $V_\mu$,
including the needed counterterms, are shown in Figure~\ref{diagrams_vv}. As done before, dimensional regularization will be employed.


\begin{figure}
	\begin{centering}
		\includegraphics[scale=0.5]{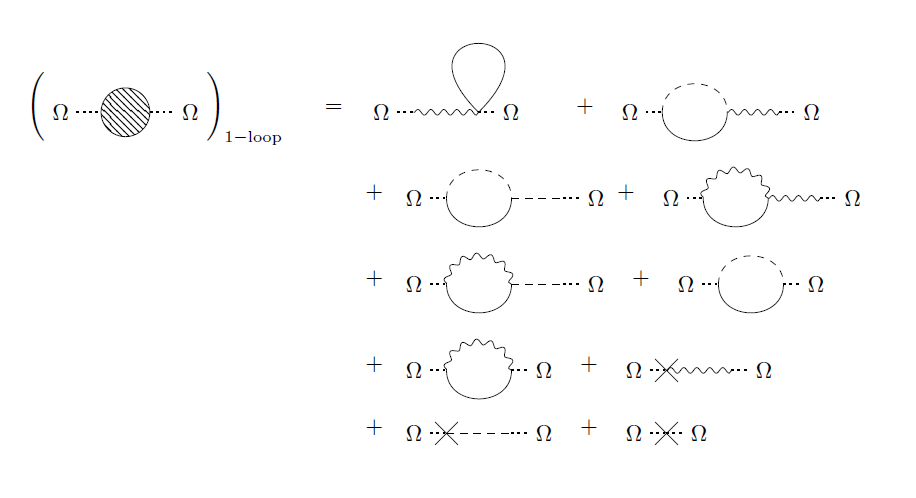}
		\par\end{centering}
	\caption{Diagrams contributing to the one-loop two-point Green's function of
		the $V_{\mu}$ operator. Curly lines refer to the gauge field, solid ones to the Higgs field and dashed ones to the Goldstone field.}
	\label{diagrams_vv}
	
\end{figure}

For further use, next to the already defined \eqref{chim}, it is helpful to also introduce
\begin{eqnarray}
\eta\left(m_{1}^{2},m_{2}^{2},p^{2}\right) & = & \int\frac{d^{d}k}{\left(2\pi\right)^{d}}\frac{1}{k^{2}+m_{1}^{2}}\frac{1}{\left(p-k\right)^{2}+m_{2}^{2}}\nonumber \\
& = & \frac{1}{\left(4\pi\right)^{\frac{d}{2}}}\Gamma\left(2-\frac{d}{2}\right)\int_{0}^{1}dx\left(p^{2}x\left(x-1\right)+m_{1}^{2}x+m_{2}^{2}\left(1-x\right)\right)^{\frac{d}{2}-2} \;. \label{etaf}
\end{eqnarray}
For the diagrams of Figure~\ref{diagrams_vv} we obtain:
\begin{eqnarray}
\widetilde{\left\langle V_{\mu}\left(p\right)V_{\nu}\left(-p\right)\right\rangle }_{\left(1\right)} & = & \left(-e\right)\left(\int\frac{d^{d}k}{\left(2\pi\right)^{d}}\frac{1}{k^{2}+m_{h}^{2}}\right)\frac{\mathcal{P}_{\mu\nu}\left(p\right)}{p^{2}+m^{2}}\left(-\frac{1}{2}ev^{2}\right)\nonumber \\
& = & \frac{1}{2}e^{2}v^{2}\frac{1}{p^{2}+m^{2}}\chi\left(m_{h}^{2}\right)\mathcal{P}_{\mu\nu}\left(p\right)\,,\label{eq:vv1}
\end{eqnarray}
\begin{eqnarray}
\widetilde{\left\langle V_{\mu}\left(p\right)V_{\nu}\left(-p\right)\right\rangle }_{\left(2\right)} & = & 2\int\frac{d^{d}k}{\left(2\pi\right)^{d}}\left[-\frac{i}{2}\left(-p_{\mu}+2k_{\mu}\right)\right]\frac{1}{k^{2}+m_{h}^{2}}\frac{1}{\left(p-k\right)^{2}}\left[ie\left(2k_{\alpha}-p_{\alpha}\right)\right]\frac{\mathcal{P}_{\alpha\nu}\left(p\right)}{p^{2}+m^{2}}\left(-\frac{1}{2}ev^{2}\right)\nonumber \\
& = & -2e^{2}v^{2}\frac{\mathcal{P}_{\mu\nu}\left(p\right)}{p^{2}+m^{2}}\frac{1}{\left(d-1\right)}\left\{ -m_{h}^{2}\eta\left(m_{h}^{2},0,p^{2}\right)\right.\nonumber \\
&  & \left.-\frac{1}{4p^{2}}\left[\left(p^{2}-m_{h}^{2}\right)^{2}\eta\left(m_{h}^{2},0,p^{2}\right)-\left(p^{2}-m_{h}^{2}\right)\chi\left(m_{h}^{2}\right)\right]\right\} \,, \label{eq:vv2}
\end{eqnarray}
\begin{eqnarray}
\widetilde{\left\langle V_{\mu}\left(p\right)V_{\nu}\left(-p\right)\right\rangle }_{\left(3\right)} & = & 2\int\frac{d^{d}k}{\left(2\pi\right)^{d}}\left[-\frac{i}{2}\left(-p_{\mu}+2k_{\mu}\right)\right]\frac{1}{k^{2}+m_{h}^{2}}\frac{1}{\left(p-k\right)^{2}}\left(-\lambda v\right)\frac{1}{p^{2}}\left(-\frac{i}{2}vp_{\nu}\right)\nonumber \\
& = & -\frac{1}{2}\lambda v^{2}\frac{1}{p^{2}}\mathcal{L}_{\mu\nu}\left(p\right)\left[m_{h}^{2}\eta\left(0,m_{h}^{2},p^{2}\right)+\chi\left(m_{h}^{2}\right)\right]\label{eq:vv3}
\end{eqnarray}
\begin{eqnarray}
\widetilde{\left\langle V_{\mu}\left(p\right)V_{\nu}\left(-p\right)\right\rangle }_{\left(4\right)} & = & 2\left(-ev\right)\int\frac{d^{d}k}{\left(2\pi\right)^{d}}\frac{\mathcal{P}_{\mu\alpha}\left(k\right)}{k^{2}+m^{2}}\frac{1}{\left(p-k\right)^{2}+m_{h}^{2}}\left(-2e^{2}v\delta_{\alpha\beta}\right)\frac{\mathcal{P}_{\beta\nu}\left(p\right)}{p^{2}+m^{2}}\left(-\frac{1}{2}ev^{2}\right)\nonumber \\
& = & -2\mathcal{P}_{\mu\nu}\left(p\right)\frac{m^{4}}{p^{2}+m^{2}}\left\{ \eta\left(m^{2},m_{h}^{2},p^{2}\right)\right.\nonumber \\
&  & +\frac{1}{\left(d-1\right)}\frac{\left(p^{2}+m_{h}^{2}\right)^{2}}{4m^{2}p^{2}}\eta\left(0,m_{h}^{2},p^{2}\right)-\frac{1}{\left(d-1\right)}\eta\left(m^{2},m_{h}^{2},p^{2}\right)\nonumber \\
&  & -\frac{1}{\left(d-1\right)4m^{2}p^{2}}\left[\left(p^{2}+m_{h}^{2}-m^{2}\right)^{2}\eta\left(m^{2},m_{h}^{2},p^{2}\right)\right.\nonumber \\
&  & \left.\left.-m^{2}\chi\left(m_{h}^{2}\right)-\left(p^{2}+m_{h}^{2}-m^{2}\right)\chi\left(m^{2}\right)\right]\right\}\,, \label{eq:vv4}
\end{eqnarray}
\begin{eqnarray}
\widetilde{\left\langle V_{\mu}\left(p\right)V_{\nu}\left(-p\right)\right\rangle }_{\left(5\right)} & = & 2\left(-ev\right)\int\frac{d^{d}k}{\left(2\pi\right)^{d}}\frac{\mathcal{P}_{\mu\alpha}\left(k\right)}{k^{2}+m^{2}}\frac{1}{\left(p-k\right)^{2}+m_{h}^{2}}\left[ie\left(-k_{\alpha}+2p_{\alpha}\right)\right]\frac{1}{p^{2}}\left(-\frac{i}{2}vp_{\nu}\right)\nonumber \\
& = & -2m^{2}\mathcal{L}_{\mu\nu}\left(p\right)\left\{ \eta\left(m^{2},m_{h}^{2},p^{2}\right)-\frac{\left(p^{2}+m_{h}^{2}\right)^{2}}{4m^{2}p^{2}}\eta\left(0,m_{h}^{2},p^{2}\right)\right.\nonumber \\
&  & +\frac{1}{4m^{2}p^{2}}\left[\left(p^{2}+m_{h}^{2}-m^{2}\right)^{2}\eta\left(m^{2},m_{h}^{2},p^{2}\right)\right.\nonumber \\
&  & \left.\left.-m^{2}\chi\left(m_{h}^{2}\right)-\left(p^{2}+m_{h}^{2}-m^{2}\right)\chi\left(m^{2}\right)\right]\right\}\,, \label{eq:vv5}
\end{eqnarray}
\begin{eqnarray}
\widetilde{\left\langle V_{\mu}\left(p\right)V_{\nu}\left(-p\right)\right\rangle }_{\left(6\right)} & = & \int\frac{d^{d}k}{\left(2\pi\right)^{d}}\left[-\frac{i}{2}\left(2k_{\mu}-p_{\mu}\right)\right]\frac{1}{k^{2}+m_{h}^{2}}\left[-\frac{i}{2}\left(-2k_{\nu}+p_{\nu}\right)\right]\frac{1}{\left(p-k\right)^{2}}\nonumber \\
& = & \mathcal{P}_{\mu\nu}\left(p\right)\frac{1}{\left(d-1\right)}\left\{ -m_{h}^{2}\eta\left(m_{h}^{2},0,p^{2}\right)\right.\nonumber \\
&  & \left.-\frac{1}{4p^{2}}\left[\left(p^{2}-m_{h}^{2}\right)^{2}\eta\left(m_{h}^{2},0,p^{2}\right)-\left(p^{2}-m_{h}^{2}\right)\chi\left(m_{h}^{2}\right)\right]\right\} \nonumber \\
&  & +\frac{1}{4}\mathcal{L}_{\mu\nu}\left(p\right)\left\{ \frac{m_{h}^{4}}{p^{2}}\eta\left(m_{h}^{2},0,p^{2}\right)+\frac{\left(p^{2}+m_{h}^{2}\right)}{p^{2}}\chi\left(m_{h}^{2}\right)\right\}\,, \label{eq:vv6}
\end{eqnarray}
\begin{eqnarray}
\widetilde{\left\langle V_{\mu}\left(p\right)V_{\nu}\left(-p\right)\right\rangle }_{\left(7\right)} & = & \int\frac{d^{d}k}{\left(2\pi\right)^{d}}\left(-ev\right)\frac{\mathcal{P}_{\mu\nu}\left(k\right)}{k^{2}+m^{2}}\left(-ev\right)\frac{1}{\left(p-k\right)^{2}+m_{h}^{2}}\nonumber \\
& = & m^{2}\mathcal{P}_{\mu\nu}\left(p\right)\left\{ \eta\left(m^{2},m_{h}^{2},p^{2}\right)+\frac{1}{\left(d-1\right)}\frac{1}{4m^{2}p^{2}}\left(p^{2}+m_{h}^{2}\right)^{2}\eta\left(0,m_{h}^{2},p^{2}\right)\right.\nonumber \\
&  & -\frac{1}{\left(d-1\right)}\eta\left(m^{2},m_{h}^{2},p^{2}\right)-\frac{1}{\left(d-1\right)}\frac{1}{4m^{2}p^{2}}\left[\left(p^{2}+m_{h}^{2}-m^{2}\right)^{2}\eta\left(m^{2},m_{h}^{2},p^{2}\right)\right.\nonumber \\
&  & \left.\left.-m^{2}\chi\left(m_{h}^{2}\right)-\left(p^{2}+m_{h}^{2}-m^{2}\right)\chi\left(m^{2}\right)\right]\right\} \nonumber \\
&  & +e^{2}v^{2}\mathcal{L}_{\mu\nu}\left(p\right)\left\{ \eta\left(m^{2},m_{h}^{2},p^{2}\right)-\frac{1}{4m^{2}p^{2}}\left(p^{2}+m_{h}^{2}\right)^{2}\eta\left(0,m_{h}^{2},p^{2}\right)\right.\nonumber \\
&  & +\frac{1}{4m^{2}p^{2}}\left[\left(p^{2}+m_{h}^{2}-m^{2}\right)^{2}\eta\left(m^{2},m_{h}^{2},p^{2}\right)\right.\nonumber \\
&  & \left.\left.-m^{2}\chi\left(m_{h}^{2}\right)-\left(p^{2}+m_{h}^{2}-m^{2}\right)\chi\left(m^{2}\right)\right]\right\}\,, \label{eq:vv7}
\end{eqnarray}
\begin{eqnarray}
\widetilde{\left\langle V_{\mu}\left(p\right)V_{\nu}\left(-p\right)\right\rangle }_{\left(8\right)} & = & 2\left(-\frac{1}{2}Z_{h}^{\left(1\right)}ev^{2}\delta_{\mu\alpha}+Z_{\Upsilon\Omega}^{\left(1\right)}p^{2}\mathcal{P}_{\mu\alpha}\left(p\right)\right)\frac{\mathcal{P}_{\alpha\nu}\left(p\right)}{p^{2}+m^{2}}\left(-\frac{1}{2}ev^{2}\right)\nonumber \\
& = & \frac{1}{2e^{2}}\left(Z_{h}^{\left(1\right)}m^{2}+Z_{A}^{\left(1\right)}p^{2}\right)\frac{m^{2}}{p^{2}+m^{2}}\mathcal{P}_{\mu\nu}\left(p\right)\,,\label{eq:vv8}
\end{eqnarray}
\begin{eqnarray}
\widetilde{\left\langle V_{\mu}\left(p\right)V_{\nu}\left(-p\right)\right\rangle }_{\left(9\right)} & = & 2\left(\frac{1}{2}Z_{h}^{\left(1\right)}vip_{\mu}\right)\frac{1}{p^{2}}\left(-\frac{1}{2}ivp_{\nu}\right)\nonumber \\
& = & \frac{1}{2}v^{2}Z_{h}^{\left(1\right)}\mathcal{L}_{\mu\nu}\left(p\right)\,,\label{eq:vv9}
\end{eqnarray}
\begin{eqnarray}
\widetilde{\left\langle V_{\mu}\left(p\right)V_{\nu}\left(-p\right)\right\rangle }_{\left(10\right)} & = & \left[-\frac{p^{2}}{4e^{2}}Z_{A}^{\left(1\right)}\mathcal{P}_{\mu\nu}\left(p\right)-\frac{v^{2}}{4}Z_{h}^{\left(1\right)}\delta_{\mu\nu}-\frac{1}{4}\chi\left(m_{h}^{2}\right)\delta_{\mu\nu}\right]\,.\label{eq:vv10}
\end{eqnarray}
Finally, summing up all contributions, eqs.~(\ref{eq:vv1})-(\ref{eq:vv10}), one gets
\begin{eqnarray}
\frac{\mathcal{P}_{\mu\nu}(p)}{(d-1)}\;\widetilde{\left\langle V_{\mu}\left(p\right)V_{\nu}\left(-p\right)\right\rangle }_{\textrm{1-loop}} & = & \frac{\left(p^{4}-m^{4}\right)}{\left(p^{2}+m^{2}\right)^{2}}\left\{ -\frac{1}{4}\chi\left(m_{h}^{2}\right)\right.\nonumber \\
&  & +\frac{1}{\left(d-1\right)}\left\{ -m_{h}^{2}\eta\left(m_{h}^{2},0,p^{2}\right)-\frac{1}{4p^{2}}\left[\left(p^{2}-m_{h}^{2}\right)^{2}\eta\left(m_{h}^{2},0,p^{2}\right)-\left(p^{2}-m_{h}^{2}\right)\chi\left(m_{h}^{2}\right)\right]\right\} \nonumber \\
&  & +m^{2}\left\{ \eta\left(m^{2},m_{h}^{2},p^{2}\right)+\frac{1}{\left(d-1\right)}\frac{\left(p^{2}+m_{h}^{2}\right)^{2}}{4p^{2}m^{2}}\eta\left(0,m_{h}^{2},p^{2}\right)\right.\nonumber \\
&  & -\frac{1}{\left(d-1\right)}\eta\left(m^{2},m_{h}^{2},p^{2}\right)-\frac{1}{\left(d-1\right)}\frac{1}{4m^{2}p^{2}}\left[\left(p^{2}+m_{h}^{2}-m^{2}\right)^{2}\eta\left(m^{2},m_{h}^{2},p^{2}\right)\right.\nonumber \\
&  & \left.\left.-m^{2}\chi\left(m_{h}^{2}\right)-\left(p^{2}+m_{h}^{2}-m^{2}\right)\chi\left(m^{2}\right)\right]\right\} \nonumber \\
&  & \left.-\frac{1}{4e^{2}}\left(Z_{A}^{\left(1\right)}p^{2}+Z_{h}^{\left(1\right)}m^{2}\right)\right\} \;, \label{eq:vvt}
\end{eqnarray}
and
\begin{eqnarray}
\mathcal{L}_{\mu\nu}(p)\widetilde{\left\langle V_{\mu}\left(p\right)V_{\nu}\left(-p\right)\right\rangle }_{\textrm{1-loop}} & = & -m^{2}\left\{ \eta\left(m^{2},m_{h}^{2},p^{2}\right)-\frac{1}{4m^{2}p^{2}}\left(p^{2}+m_{h}^{2}\right)^{2}\eta\left(0,m_{h}^{2},p^{2}\right)\right.\nonumber \\
&  & +\frac{1}{4m^{2}p^{2}}\left[\left(p^{2}+m_{h}^{2}-m^{2}\right)^{2}\eta\left(m^{2},m_{h}^{2},p^{2}\right)\right.\nonumber \\
&  & \left.\left.-m^{2}\chi\left(m_{h}^{2}\right)-\left(p^{2}+m_{h}^{2}-m^{2}\right)\chi\left(m^{2}\right)\right]\right\} \nonumber \\
&  & -\frac{1}{4}\frac{m_{h}^{2}}{p^{2}}\left[m_{h}^{2}\eta\left(0,m_{h}^{2},p^{2}\right)+\chi\left(m_{h}^{2}\right)\right]+\frac{1}{4}v^{2}Z_{h}^{\left(1\right)}.\label{eq:vvl}
\end{eqnarray}

\subsection{Evaluation of $\left\langle A_{\mu}\left(p\right)A_{\nu}\left(-p\right)\right\rangle $}

\begin{figure}
	\begin{centering}
		\includegraphics[scale=0.6]{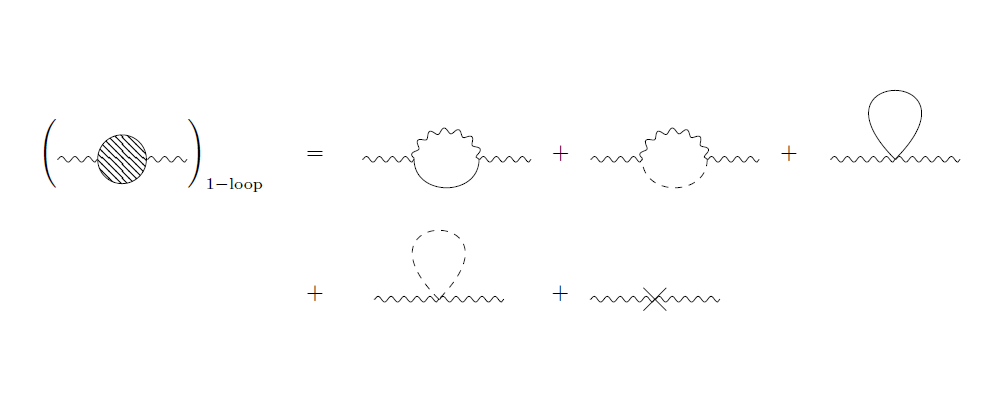}
		\par\end{centering}
	\caption{Diagrams contributing to the one-loop two-point Green's function of
		the Abelian gauge field $A_{\mu}$. Curly lines refer to the gauge field, solid ones to the Higgs field and dashed ones to the Goldstone field.}
	\label{diagrams_aa}
	
\end{figure}

At one-loop order, for the two-point function of the gauge field $A_{\mu}$ we have
the diagrams depicted  in Figure~\ref{diagrams_aa}:
\begin{eqnarray}
\left\langle A_{\mu}\left(p\right)A_{\nu}\left(-p\right)\right\rangle_{\left(1\right)} & = & \frac{\mathcal{P}_{\mu\alpha}\left(p\right)}{p^{2}+m^{2}}\frac{\mathcal{P}_{\nu\beta}\left(p\right)}{p^{2}+m^{2}}\left(-2e^{2}v\delta_{\alpha\rho}\right)\left(-2e^{2}v\delta_{\beta\lambda}\right)\int\frac{d^{d}k}{\left(2\pi\right)^{d}}\frac{P_{\rho\lambda}\left(k\right)}{k^{2}+m^{2}}\frac{1}{\left(p-k\right)^{2}+m_{h}^{2}}\nonumber \\
&  & =4e^{2}m^{2}\frac{\mathcal{P}_{\mu\nu}\left(p\right)}{\left(p^{2}+m^{2}\right)^{2}}\nonumber \\
&  & \left\{ \eta\left(m^{2},m_{h}^{2},p^{2}\right)+\frac{1}{\left(d-1\right)}\frac{\left(p^{2}+m_{h}^{2}\right)^{2}}{4m^{2}p^{2}}\eta\left(0,m_{h}^{2},p^{2}\right)-\frac{1}{\left(d-1\right)}\eta\left(m^{2},m_{h}^{2},p^{2}\right)\right.\nonumber \\
&  & -\frac{1}{\left(d-1\right)}\frac{1}{4m^{2}p^{2}}\left[\left(p^{2}+m_{h}^{2}-m^{2}\right)^{2}\eta\left(m^{2},m_{h}^{2},p^{2}\right)\right.\nonumber \\
&  & \left.\left.-m^{2}\chi\left(m_{h}^{2}\right)-\left(p^{2}+m_{h}^{2}-m^{2}\right)\chi\left(m^{2}\right)\right]\right\}\,, \label{eq:aa1}
\end{eqnarray}
\begin{eqnarray}
\left\langle A_{\mu}\left(p\right)A_{\nu}\left(-p\right)\right\rangle _{\left(2\right)} & = & \frac{\mathcal{P}_{\mu\alpha}\left(p\right)}{p^{2}+m^{2}}\frac{\mathcal{P}_{\nu\beta}\left(p\right)}{p^{2}+m^{2}}\int\frac{d^{d}k}{\left(2\pi\right)^{d}}ie\left(-2k_{\alpha}+p_{\alpha}\right)\frac{1}{k^{2}+m_{h}^{2}}ie\left(2k_{\beta}-p_{\beta}\right)\frac{1}{\left(p-k\right)^{2}}\nonumber \\
& = & 4e^{2}\frac{\mathcal{P}_{\mu\nu}\left(p\right)}{\left(p^{2}+m^{2}\right)^{2}}\frac{1}{\left(d-1\right)}\left\{ -m_{h}^{2}\eta\left(m_{h}^{2},0,p^{2}\right)\right.\nonumber \\
&  & \left.-\frac{1}{4p^{2}}\left[\left(p^{2}-m_{h}^{2}\right)^{2}\eta\left(m_{h}^{2},0,p^{2}\right)-\left(p^{2}-m_{h}^{2}\right)\chi\left(m_{h}^{2}\right)\right]\right\}\,, \label{eq:aa2}
\end{eqnarray}
\begin{eqnarray}
\left\langle A_{\mu}\left(p\right)A_{\nu}\left(-p\right)\right\rangle _{\left(3\right)} & = & \frac{1}{2}\frac{\mathcal{P}_{\mu\alpha}\left(p\right)}{p^{2}+m^{2}}\frac{\mathcal{P}_{\nu\beta}\left(p\right)}{p^{2}+m^{2}}\left(-2e^{2}\delta_{\alpha\beta}\right)\int\frac{d^{d}k}{\left(2\pi\right)^{d}}\frac{1}{k^{2}+m_{h}^{2}}\nonumber \\
& = & \frac{\mathcal{P}_{\mu\nu}\left(p\right)}{\left(p^{2}+m^{2}\right)^{2}}\left(-e^{2}\chi\left(m_{h}^{2}\right)\right)\label{eq:aa3}\,,
\end{eqnarray}
\begin{eqnarray}
\left\langle A_{\mu}\left(p\right)A_{\nu}\left(-p\right)\right\rangle _{\left(4\right)} & = & 0\label{eq:aa4}\,,
\end{eqnarray}
\begin{eqnarray}
\left\langle A_{\mu}\left(p\right)A_{\nu}\left(-p\right)\right\rangle _{\left(5\right)} & = & -\frac{\mathcal{P}_{\mu\nu}\left(p\right)}{\left(p^{2}+m^{2}\right)^{2}}\left(Z_{A}^{\left(1\right)}p^{2}+Z_{h}^{\left(1\right)}m^{2}\right).\label{eq:aa5}
\end{eqnarray}
Summing up all contributions, eqs.~(\ref{eq:aa1})-(\ref{eq:aa5}),
we find that
\begin{eqnarray}
\frac{\mathcal{P}_{\mu\nu}\left(p\right)}{(d-1)}\left\langle A_{\mu}\left(p\right)A_{\nu}\left(-p\right)\right\rangle _{\textrm{1-loop}} & = & \frac{4e^{2}m^{2}}{\left(p^{2}+m^{2}\right)^{2}}\left\{ \eta\left(m^{2},m_{h}^{2},p^{2}\right)+\frac{1}{\left(d-1\right)}\frac{\left(p^{2}+m_{h}^{2}\right)^{2}}{4m^{2}p^{2}}\eta\left(0,m_{h}^{2},p^{2}\right)\right.\nonumber \\
&  & -\frac{1}{\left(d-1\right)}\eta\left(m^{2},m_{h}^{2},p^{2}\right)-\frac{1}{\left(d-1\right)}\frac{1}{4m^{2}p^{2}}\left[\left(p^{2}+m_{h}^{2}-m^{2}\right)^{2}\eta\left(m^{2},m_{h}^{2},p^{2}\right)\right.\nonumber \\
&  & \left.\left.-m^{2}\chi\left(m_{h}^{2}\right)-\left(p^{2}+m_{h}^{2}-m^{2}\right)\chi\left(m^{2}\right)\right]\right\} \nonumber \\
&  & +\frac{4e^{2}}{\left(p^{2}+m^{2}\right)^{2}}\frac{1}{\left(d-1\right)}\left\{ -m_{h}^{2}\eta\left(m_{h}^{2},0,p^{2}\right)\right.\nonumber \\
&  & \left.-\frac{1}{4p^{2}}\left[\left(p^{2}-m_{h}^{2}\right)^{2}\eta\left(m_{h}^{2},0,p^{2}\right)-\left(p^{2}-m_{h}^{2}\right)\chi\left(m_{h}^{2}\right)\right]\right\} \nonumber \\
&  & -\frac{1}{\left(p^{2}+m^{2}\right)^{2}}e^{2}\chi\left(m_{h}^{2}\right)\nonumber \\
&  & -\frac{1}{\left(p^{2}+m^{2}\right)^{2}}\left(Z_{A}^{\left(1\right)}p^{2}+Z_{h}^{\left(1\right)}m^{2}\right).\label{eq:aafinal}
\end{eqnarray}
Comparing now the two expressions (\ref{eq:vvt}) and (\ref{eq:aafinal}), one immediately realizes that the Ward identity (\ref{eq:transversal_result-1-1}) is fulfilled at the one-loop order.

\subsection{Evaluation of $\left\langle \rho\left(p\right)\rho\left(-p\right)\right\rangle $}

\begin{figure}
	\begin{centering}
		\includegraphics[scale=0.6]{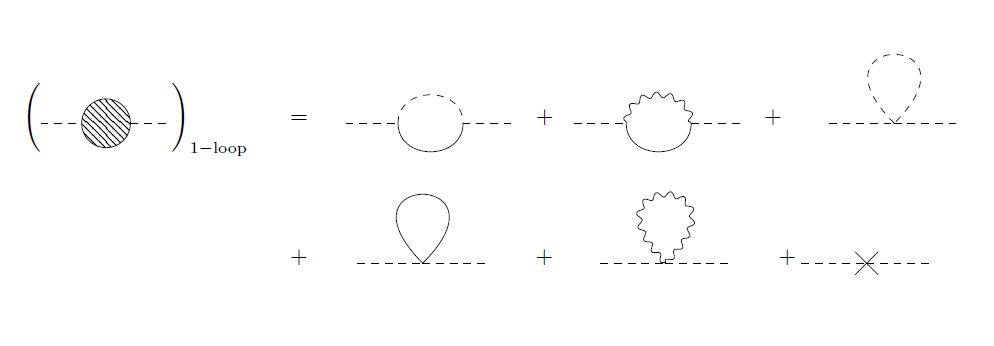}
		\par\end{centering}
	\caption{Feynman diagrams contributing to the two-point Green's function of
		the Goldstone field $\rho$. Curly lines refer to the gauge field, solid ones to the Higgs field and dashed ones to the Goldstone field.}
	\label{diagrams_rr}
	
\end{figure}

At one-loop order, for the two-point function of the Goldstone field $\rho$ we have the
diagrams shown in Figure \ref{diagrams_rr}. They are given by:

\begin{eqnarray}
\left\langle \rho\left(p\right)\rho\left(-p\right)\right\rangle _{\left(1\right)} & = & \frac{1}{p^{4}}\left(-\lambda v\right)^{2}\int\frac{d^{d}k}{\left(2\pi\right)^{d}}\frac{1}{k^{2}}\frac{1}{\left(p-k\right)^{2}+m_{h}^{2}}\nonumber \\
& = & \frac{1}{p^{4}}\left(-\lambda v\right)^{2}\eta\left(0,m_{h}^{2},p^{2}\right)\,,\label{eq:rhorho1}
\end{eqnarray}
\begin{center}
	\begin{eqnarray}
	\left\langle \rho\left(p\right)\rho\left(-p\right)\right\rangle _{\left(2\right)} & = & \frac{1}{p^{4}}\int\frac{d^{d}k}{\left(2\pi\right)^{d}}\frac{\mathcal{P}_{\mu\nu}\left(k\right)}{k^{2}+m^{2}}ie\left(-2p_{\mu}+k_{\mu}\right)\frac{1}{\left(p-k\right)^{2}+m_{h}^{2}}ie\left(2p_{\nu}-k_{\nu}\right)\nonumber \\
	& = & 4e^{2}\frac{1}{p^{2}}\left\{ \eta\left(m^{2},m_{h}^{2},p^{2}\right)-\frac{1}{m^{2}}\frac{1}{4p^{2}}\left(p^{2}+m_{h}^{2}\right)^{2}\eta\left(0,m_{h}^{2},p^{2}\right)\right.\nonumber \\
	&  & +\frac{1}{m^{2}}\frac{1}{4p^{2}}\left[\left(p^{2}+m_{h}^{2}-m^{2}\right)^{2}\eta\left(m^{2},m_{h}^{2},p^{2}\right)\right.\nonumber \\
	&  & \left.\left.-m^{2}\chi\left(m_{h}^{2}\right)-\left(p^{2}+m_{h}^{2}-m^{2}\right)\chi\left(m^{2}\right)\right]\right\}\,, \label{eq:rhorho2}
	\end{eqnarray}
	\par\end{center}
\begin{eqnarray*}
	\left\langle \rho\left(p\right)\rho\left(-p\right)\right\rangle _{\left(3\right)} & = & 0\,,
\end{eqnarray*}
\begin{center}
	\begin{eqnarray}
	\left\langle \rho\left(p\right)\rho\left(-p\right)\right\rangle _{\left(4\right)} & = & \frac{1}{2}\frac{1}{p^{4}}\left(-\lambda\right)\int\frac{d^{d}k}{\left(2\pi\right)^{d}}\frac{1}{k^{2}+m_{h}^{2}}\nonumber \\
	& = & -\frac{1}{2}\frac{1}{p^{4}}\lambda\chi\left(m_{h}^{2}\right)\label{eq:rhorho4}\,,
	\end{eqnarray}
	\par\end{center}

\begin{eqnarray}
\left\langle \rho\left(p\right)\rho\left(-p\right)\right\rangle _{\left(5\right)} & = & \frac{1}{2}\frac{1}{p^{4}}\left(-2e^{2}\delta_{\mu\nu}\right)\int\frac{d^{d}k}{\left(2\pi\right)^{d}}\frac{\mathcal{P}_{\mu\nu}\left(k\right)}{k^{2}+m^{2}}\nonumber \\
& = & -\frac{1}{p^{4}}e^{2}\left(d-1\right)\chi\left(m^{2}\right)\,,\label{eq:rhorho5}
\end{eqnarray}
\begin{eqnarray}
\left\langle \rho\left(p\right)\rho\left(-p\right)\right\rangle _{\left(6\right)} & = & -\frac{1}{p^{4}}\left(Z_{h}^{\left(1\right)}p^{2}+\delta\sigma^{\left(1\right)}v^{2}\right).\label{eq:rhorho6}
\end{eqnarray}
Summing up all contributions, eqs.~(\ref{eq:rhorho1})-(\ref{eq:rhorho6}),
we get
\begin{eqnarray}
\left\langle \rho\left(p\right)\rho\left(-p\right)\right\rangle _{\textrm{1-loop}} & = & 4e^{2}\frac{1}{p^{2}}\left\{ \eta\left(m^{2},m_{h}^{2},p^{2}\right)-\frac{1}{m^{2}}\frac{1}{4p^{2}}\left(p^{2}+m_{h}^{2}\right)^{2}\eta\left(0,m_{h}^{2},p^{2}\right)\right.\nonumber \\
&  & +\frac{1}{m^{2}}\frac{1}{4p^{2}}\left[\left(p^{2}+m_{h}^{2}-m^{2}\right)^{2}\eta\left(m^{2},m_{h}^{2},p^{2}\right)\right.\nonumber \\
&  & \left.\left.-m^{2}\chi\left(m_{h}^{2}\right)-\left(p^{2}+m_{h}^{2}-m^{2}\right)\chi\left(m^{2}\right)\right]\right\} \nonumber \\
&  & +\frac{\lambda}{p^{4}}\left[m_{h}^{2}\eta\left(0,m_{h}^{2},p^{2}\right)+\chi\left(m_{h}^{2}\right)\right]-\frac{1}{p^{2}}Z_{h}^{\left(1\right)} \;.\label{eq:rhorhofinal}
\end{eqnarray}
Again, the direct comparison of the two equations (\ref{eq:vvl}) and (\ref{eq:rhorhofinal}) shows that the Ward identity (\ref{eq:longitudinal_result-1-1}) for the longitudinal sector is also fulfilled at one-loop order.


\begin{thebibliography}{99}


\bibitem{Dudal:2019aew}
  D.~Dudal, D.~M.~van Egmond, M.~S.~Guimaraes, O.~Holanda, B.~W.~Mintz, L.~F.~Palhares, G.~Peruzzo and S.~P.~Sorella,
  ``Some remarks on the spectral functions of the Abelian Higgs Model,''
  Phys.\ Rev.\ D {\bf 100}, no. 6, 065009 (2019)
  [arXiv:1905.10422 [hep-th]].

\bibitem{Dudal:2019pyg}
  D.~Dudal, D.~M.~van Egmond, M.~S.~Guimaraes, O.~Holanda, L.~F.~Palhares, G.~Peruzzo and S.~P.~Sorella,
  ``Gauge-invariant spectral description of the $U(1)$ Higgs model from local composite operators,''
  JHEP {\bf 2002}, 188 (2020)
  [arXiv:1912.11390 [hep-th]].

\bibitem{Capri:2020ppe}
M.~A.~L.~Capri, I.~F.~Justo, L.~F.~Palhares, G.~Peruzzo and S.~P.~Sorella,
``Study of the renormalization of BRST invariant local composite operators in the $U(1)$ Higgs model,''
Phys. Rev. D \textbf{102} no.3, 033003 (2020)
[arXiv:2007.01770 [hep-th]].


 \bibitem{hooft2012nonperturbative}

G.~'t Hooft, A.~Jaffe, G.~Mack, P.~K.~Mitter and R.~Stora,
``Nonperturbative quantum field theory,''
NATO Sci. Ser. B \textbf{185} (1988), pp.1-603.

\bibitem{Frohlich:1980gj}
  J.~Frohlich, G.~Morchio and F.~Strocchi,
 ``Higgs Phenomenon Without A Symmetry Breaking Order Parameter,''
  Phys.\ Lett.\  {\bf 97B}, 249 (1980).

\bibitem{Frohlich:1981yi}
  J.~Frohlich, G.~Morchio and F.~Strocchi,
  ``Higgs Phenomenon Without Symmetry Breaking Order Parameter,''
  Nucl.\ Phys.\ B {\bf 190}, 553 (1981).



\bibitem{Maas:2019nso}
  A.~Maas,
  ``Brout-Englert-Higgs physics: From foundations to phenomenology,''
  Prog.\ Part.\ Nucl.\ Phys.\  {\bf 106}, 132 (2019)
  [arXiv:1712.04721 [hep-ph]].

\bibitem{Maas:2017xzh}
  A.~Maas, R.~Sondenheimer and P.~T\"orek,
  ``On the observable spectrum of theories with a Brout-Englert-Higgs effect,''
  Annals Phys.\  {\bf 402}, 18 (2019)
  [arXiv:1709.07477 [hep-ph]].

\bibitem{Sondenheimer:2019idq}
 R.~Sondenheimer,
``Analytical relations for the bound state spectrum of gauge theories with a Brout-Englert-Higgs mechanism,''
Phys. Rev. D \textbf{101} no.5, 056006 (2020)
[arXiv:1912.08680 [hep-th]].


\bibitem{Itzykson:1980rh}
 C.~Itzykson and J.~B.~Zuber,
``Quantum Field Theory,'' New York, USA: McGraw-Hill (1980).


\bibitem{Kraus:1995jk}
  E.~Kraus and K.~Sibold,
  ``Rigid invariance as derived from BRS invariance: The Abelian Higgs model,''
  Z.\ Phys.\ C {\bf 68}, 331 (1995)
  [hep-th/9503140].

\bibitem{Haussling:1996rq}
  R.~Haussling and E.~Kraus,
  ``Gauge parameter dependence and gauge invariance in the Abelian Higgs model,''
  Z.\ Phys.\ C {\bf 75}, 739 (1997).
  [hep-th/9608160].

\bibitem{Collins:2005nj}
J.~C.~Collins, A.~V.~Manohar and M.~B.~Wise,
``Renormalization of the vector current in QED,''
Phys. Rev. D \textbf{73}, 105019 (2006).
[arXiv:hep-th/0512187 [hep-th]].


\bibitem{Becchi:1974md}
C.~Becchi, A.~Rouet and R.~Stora,
``Renormalization of the Abelian Higgs-Kibble Model,''
Commun.\ Math.\ Phys.\  \textbf{42}, 127-162 (1975).

\bibitem{Becchi:1974xu}
C.~Becchi, A.~Rouet and R.~Stora,
``The Abelian Higgs-Kibble Model. Unitarity of the $S$ Operator,''
Phys.\ Lett.\ B \textbf{52}, 344-346 (1974).




\bibitem{Bergere:1975tr}
M.~C.~Bergere and Y.~M.~P.~Lam,
``Equivalence Theorem and Faddeev-Popov Ghosts,''
Phys. Rev. D \textbf{13}, 3247-3255 (1976).

\bibitem{Haag:1958vt}
R.~Haag,
``Quantum field theories with composite particles and asymptotic conditions,''
Phys. Rev. \textbf{112}, 669-673 (1958).

\bibitem{Kamefuchi:1961sb}
S.~Kamefuchi, L.~O'Raifeartaigh and A.~Salam,
``Change of variables and equivalence theorems in quantum field theories,''
Nucl. Phys. \textbf{28}, 529-549 (1961).

\bibitem{Lam:1973qa}
Y.~M.~P.~Lam,
``Equivalence theorem on Bogolyubov-Parasiuk-Hepp-Zimmermann renormalized Lagrangian field theories,''
Phys. Rev. D \textbf{7}, 2943-2949 (1973).

\bibitem{Blasi:1998ph}
A.~Blasi, N.~Maggiore, S.~P.~Sorella and L.~C.~Q.~Vilar,
``Renormalizability of nonrenormalizable field theories,''
Phys. Rev. D \textbf{59}, 121701 (1999)
[arXiv:hep-th/9812040 [hep-th]].



\bibitem{Higgs:1964pj}
P.~W.~Higgs,
``Broken Symmetries and the Masses of Gauge Bosons,''
Phys.\ Rev.\ Lett.\  \textbf{13}, 508-509 (1964).


\bibitem{Higgs:1964ia}
P.~W.~Higgs,
``Broken symmetries, massless particles and gauge fields,''
Phys.\ Lett.\  \textbf{12}, 132-133 (1964).

\bibitem{Englert:1964et}
F.~Englert and R.~Brout,
``Broken Symmetry and the Mass of Gauge Vector Mesons,''
Phys.\ Rev.\ Lett.\  \textbf{13}, 321-323 (1964).

\bibitem{Guralnik:1964eu}
G.~Guralnik, C.~Hagen and T.~Kibble,
``Global Conservation Laws and Massless Particles,''
Phys.\ Rev.\ Lett.\  \textbf{13}, 585-587 (1964).

\bibitem{Clark:1974eq}
T.~E.~Clark,
``The Abelian Higgs Model in the Landau Gauge,''
Nucl.\ Phys.\ B \textbf{90}, 484-500 (1975).


\bibitem{Piguet:1995er}
O.~Piguet and S.~P.~Sorella,
``Algebraic renormalization: Perturbative renormalization, symmetries and anomalies,''
Lect. Notes Phys. Monogr. \textbf{28}, 1-134 (1995).

\bibitem{Collins:1984xc}
 J.~C.~Collins,
``Renormalization: An Introduction to Renormalization, The Renormalization Group, and the Operator Product Expansion,'' Cambridge Monographs on Mathematical Physics 26, Cambridge University Press (1986).

\bibitem{weinberg}
 S.~Weinberg, ``The Quantum Theory of Fields. Vol.~2: Modern applications'', Cambridge University Press (2013).



\bibitem{Peskin:1995ev}
  M.~E.~Peskin and D.~V.~Schroeder,
  ``An Introduction to quantum field theory,'' Addison-Wesley,
Reading, USA, 1995.







\bibitem{Delduc:1996yh}
 F.~Delduc, N.~Maggiore, O.~Piguet and S.~Wolf,
``Note on constrained cohomology,''
Phys. Lett. B \textbf{385}, 132-138 (1996)
[arXiv:hep-th/9605158 [hep-th]].





\bibitem{Ouvry:1988mm}
S.~Ouvry, R.~Stora and P.~van Baal,
``On the Algebraic Characterization of Witten's Topological Yang-Mills Theory,''
Phys. Lett. B \textbf{220}, 159-163(1989).

\bibitem{Stora:1996yc}
R.~Stora,
``Exercises in equivariant cohomology and topological theories,''
[arXiv:hep-th/9611116 [hep-th]].

\bibitem{prep} Work in preparation.

\bibitem{Nielsen:1973cs}
 H.~B.~Nielsen and P.~Olesen,
``Vortex Line Models for Dual Strings,''
Nucl. Phys. B \textbf{61}, 45-61 (1973).

\bibitem{Coleman:1973jx}
 S.~R.~Coleman and E.~J.~Weinberg,
``Radiative Corrections as the Origin of Spontaneous Symmetry Breaking,''
Phys. Rev. D \textbf{7}, 1888-1910 (1973).

\bibitem{Knecht:2001cc}
 K.~Knecht and H.~Verschelde,
``A New start for local composite operators,''
Phys. Rev. D \textbf{64}, 085006 (2001)
[arXiv:hep-th/0104007 [hep-th]].


\end{thebibliography}
\end{document}